# The nitrogen operating space of world food production

Souhil Harchaoui* and Petros Chatzimpiros*
Laboratoire Interdisciplinaire des Énergies de Demain (LIED, UMR 8236), Université de Paris, 75013, Paris, France
* Correspondence: sharchaoui@yahoo.fr, petros.chatzimpiros@univ-paris-diderot.fr

**Agriculture operates within a global ecosystem for which planetary boundaries have recently been defined. Efficiency in nitrogen use is essential for maximizing the benefits of agriculture for humanity and minimizing adverse socio-ecological impacts. The set of variables that support global system efficiency also determine the food production boundaries of agriculture, which govern the maximum supportable human population. Food production boundaries, nitrogen loss and nitrogen self-sufficiency are combined here into the nitrogen operating space of world food production. We position world regions and the world trajectory (1961-2013) within the nitrogen operating space and show that the maximum supportable human population ranges from 6 to almost 17 billion people according to the share of grain used as feed and the nitrogen fertilization regime. All UN population projections for the 21$^{st}$ century can only be conditionally achieved. We discuss the growth rate requirements in production and efficiency to meet food production boundaries and the nitrogen planetary boundary by 2050.**

Agriculture is the largest user of land on the planet and is responsible for a large share of global environmental impacts [1,2]. Global agricultural production and pollution were historically circumscribed by low nitrogen (N) fixation until the mastery of fossil-fueled N industrial fertilizers production by the early 20$^{th}$ century [3]. Inputs from industrial N fertilizers ($N_{ind}$) today exceed atmospheric N deposition ($N_{atm}$) and biological nitrogen fixation (BNF) [4], heavily disconnecting agricultural production from N self-sufficiency. $N_{ind}$ is estimated to have underpinned a net doubling in global human population [5], while also being the largest contributor to the alteration of the global N cycle causing planetary ecosystem disturbance [6–8].
Over the last decades, agricultural system analysts have set the grounds for understanding global [7] and regional [9] N balances and quantifying the impacts of diets, production practices and feed trade on food production and the N cycle [5,10–13]. More recently, environmental impacts as driven by food demand have been synthesized within a context of planetary boundaries [11,14] to assess food system intervention measures on sustainability goals [15,16]. Planetary boundary analysis is a powerful concept to quantify human-driven overshooting on life-supporting resources and demonstrates that the N planetary boundary is the most heavily transgressed.
The notion of boundary also applies to the feeding capacity of agriculture. Food production boundaries are governed by biophysical and sociotechnical factors and set quantitative limits to supportable human population [17,18]. Food production boundaries also underpin maximum planetary N pollution as a function of diets and global system efficiency. Here, we integrate food production boundaries and the N planetary boundary into the N operating space of world food production. We use a model of N budgets integrating the key variables that drive food yield, N loss and N self-sufficiency to define the N operating space of different fertilization options and position the world regions and the historical world trajectory (1961-2013) within this space.
We provide in figure 1 a graphical representation of the N flows in the model. The model variables are listed in Table 1 together with the minimum and maximum observed values in world regions and the world average. The model distinguishes between two land categories to differentiate land in support of human-edible primary production (major crops land) from land in support of human-inedible production (N fixing land) which is only convertible to food by livestock with efficiency far below one. The ratio of N fixing to total agricultural land is the N fixing ratio (see Methods and Supplementary section 1 and Fig. S1) and amounts to 74% at the global scale today after a slight decrease since 1961 and is projected to remain unchanged to 2030 [19]. Based on BNF data from literature aggregated per land category (see Supplementary section 2 and Table S1), the N fixing land accounts for 85% of the world's agricultural BNF. We highlight the paramount role of the N fixing ratio on food yield and N self-sufficiency and show that all population projections to 2050 and 2100 can only be conditionally achieved. We find that maximum supportable human population on current agricultural land varies from 6 to almost 17 billion people according to the share of grain used as feed (feed-grain) and the N fertilization regime. With the current feed-grain, the medium UN population projection to 2050 (9.8 billion people) hits the food production boundary. We discuss the growth rates required in production and efficiency to respond to this challenge, while bringing the N operating space within the N planetary boundary.

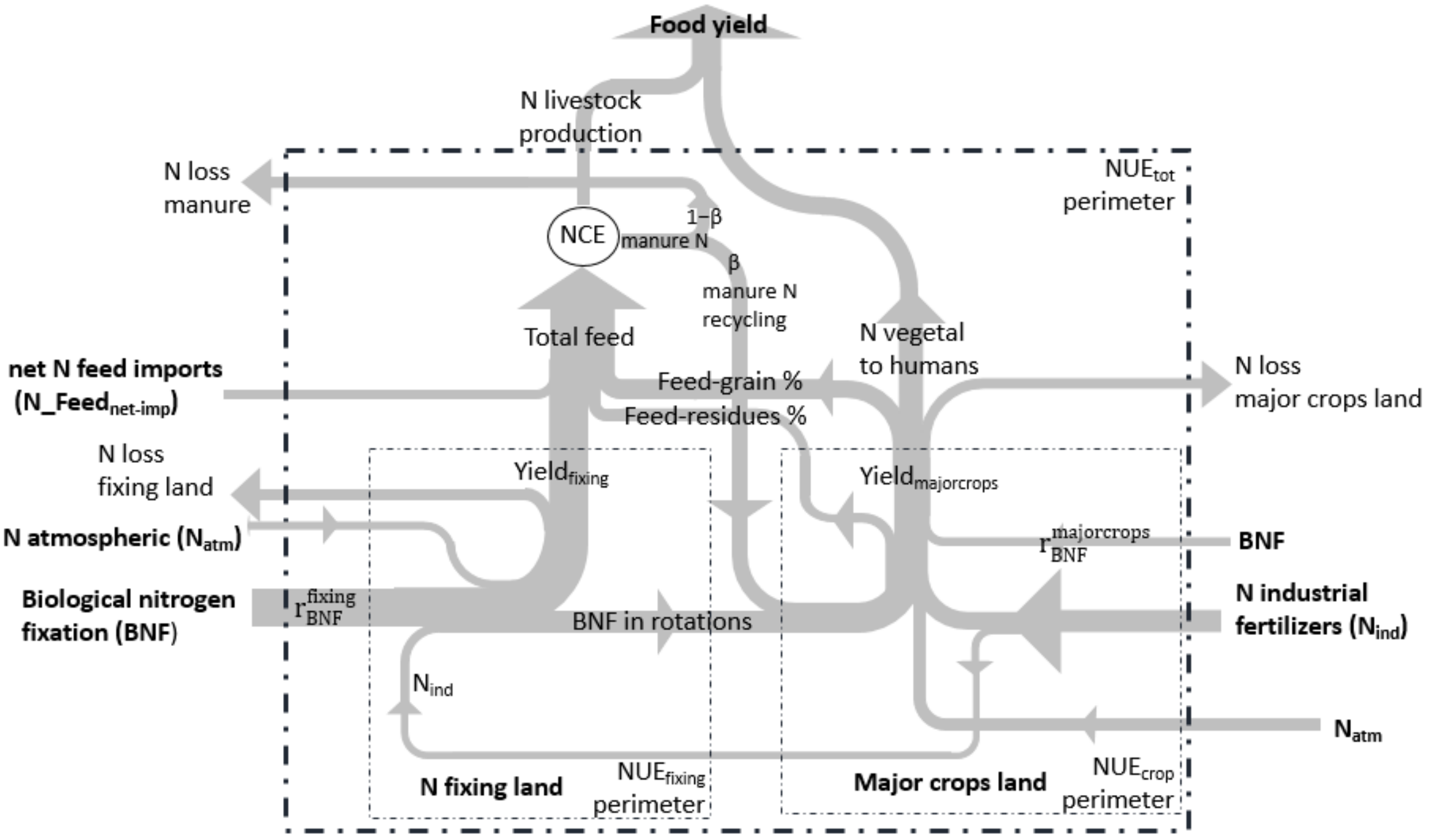

**Figure 1| Schematic representation of the modeling framework of the N operating space of food production**. The model distinguishes N fixing land from major crops land and connects total N input to food yield and environmental loss (see Supplementary section 3). Total N input ($N_{tot}$) is the sum of N from biological nitrogen fixation (BNF), atmospheric deposition ($N_{atm}$), industrial N fertilizer ($N_{ind}$) and net feed trade ($N_Feed_{net-imp}$). The food yield is primary production extracted from N fixing land ($Yield_{fixing}$) and major crops land ($Yield_{majorcrops}$) plus livestock production minus total feed (including crop residues). Total system N use efficiency ($NUE_{tot}$) is the ratio of food yield to total N input and breaks-down into efficiency in N fixing land ($NUE_{fixing}$), major crops land ($NUE_{crop}$), livestock production (NCE) and manure N recycling to major crops land (β). Total N loss is the difference between total N input and the food yield.

**Table 1| Structural model variables and numerical values used in the calculation of the productivity frontier**, based on regional minimum and maximum observed values and the world average. Sources: (1) calculated based on FAOSTAT [20] see Supplementary Table S2 (2) based on *Smil* [7] (3) based on *Billen et al.* [21] and *Bouwman et al.* [22] (4) maximum NCE based on *Smil* [23], see Supplementary section 4 (5) Not defined due to high uncertainty on grassland productivity [24] (6) calculated based on *Anglade et al.* [25] (7) estimated to balance N feed requirement considering NCE of 10% and a reasonable share of feed residues recovery close to *Bodirsky et al.* [10] and *Smil* [7] (8) calculated based on *Zhang et al.* [26] for all cropland (8a) and the major crops land as defined in this study (8b) (9) based on *Smil* [27] (10) calculated based on FAOSTAT, see Methods and Supplementary Table S3.

| Variables | Unit | Regional min-max | World current | Productivity frontier |
|---|---|---|---|---|
| N fixing ratio | % | 16-96[(1)] | 74[(1)] | Floating |
| $N_{atm}$ | kgN·ha$^{-1}$ | N.D | 5[(2)] | 5 |
| Livestock NCE | % | 3-25[(3)] | 10[(3)] | 20[(4)] |
| Manure recycling rate % | % | N.D. | 24[(2)] | Range [0, 70] |
| N fixing land | | | | |
| BNF rate ($r_{BNF}^{fixing}$) | kgN·ha$^{-1}$ | 5-38[(1)] | 15[(1)] | 35 |
| $NHI_{fixing}$ | % | N.D[(5)] | N.D[(5)] | 80[(6)] |
| $NUE_{fixing}$ | % | N.D[(5)] | N.D[(5)] | 90 |
| $Yield_{fixing}$ | kgN·ha$^{-1}$ | N.D[(5)] | 17[(7)] | 33 |
| Major crops land | | | | |
| BNF rate ($r_{BNF}^{majorcrops}$) | kgN·ha$^{-1}$ | 0.6-15[(1)] | 7.5[(1)] | 15 |
| $NUE_{crop}$ | % | 25-78[(8a)] | 38.5[(8b)] | 80 |
| $Yield_{majorcrops}$ | kgN·ha$^{-1}$ | 14-118[(1)] | 47[(1)] | 130 |
| Harvest Index | % | N.D. | 42[(9)] | 60[(9)] |
| Feed-grain | % | 15-73[(10)] | 46[(10)] | 0 |
| Feed-residues | % | N.D. | 35[(7)] | 35 |
| N industrial | kgN·ha$^{-1}$ | 3-186[(1)] | 88[(1)] | Floating |

## The productivity frontier of agriculture

The productivity frontier in fig 2 is the maximum food yield per agricultural hectare as a function of the N fixing ratio. It underlines the conflicting relationship between N self-sufficiency and agricultural productivity. The higher the N self-sufficiency, the higher the dependency of food yield on manure N recycling. We evaluate between four and ten the food yield gaps between high and low N self-sufficient systems depending on the share of manure N recycling. The composition of the food yield, N loss and N input per source in the productivity frontier is given in the Supplementary section 5 and Fig. S2. The share of livestock production increases with N self-sufficiency and highlights the inability to sustain vegan diets without industrial N fertilizers. The prevailing narrative in favor of veganism as a means to maximize food yield is meaningful only for animal proteins derived from grain-fed livestock, mainly monogastrics, but false for livestock fed on N fixing land, mainly ruminants, despite ruminant's lower NCE than monogastrics. If the N fixing ratio, which governs three-quarters of global agricultural land, is ignored, food security analysis misses its objective.

It is arguable that the relationship between livestock production and N self-sufficiency may deviate in systems with high cropland allocation to pulses and rice, which both support BNF on major crops land. However, pulses and rice are quite limited at the global scale and a large-scale increase is hard to imagine. Pulses account for only 6% of the world's cropland and for 5% of the human proteins intake [20]. In Southern Asia, which is both the largest producer and consumer of pulses, pulses account for 10% of proteins in human diets which is close to healthy diet recommendations [28]. The resulting $r_{BNF}^{majorcrops}$ in that sub-region is about 12 kgN·ha$^{-1}$, which is 5 kgN·ha$^{-1}$ above the global average (see Supplementary Table S2). Rice, which accounts for 10% of world cropland, is heavily constrained by land suitability and fresh-water requirements in support of flooding [29]. Asia gathers 88% of global rice production, a third of which (35%) is located in South-Eastern Asia [20]. In this latter sub-region, rice occupies more than 40% of total cropland resulting in average $r_{BNF}^{majorcrops}$ of 15 kgN·ha$^{-1}$, which is almost twice as high as the global average (Table 1). This highest world-region $r_{BNF}^{majorcrops}$ value is used in the productivity frontier even though its implementation at the global scale is unlikely from both production and dietary standpoints.

*Food yield gaps with world regions*

The food yield and N self-sufficiency of world regions in year 2013 are projected on the productivity frontier according to the N fixing ratio. Regional food yield gaps can be reduced by closing crop yield gaps and by reducing feed-grain. Crop yield gaps among regions are in part incompressible due to biophysical regional inequity (Table S2), whereas feed-grain use is entirely a human choice. We outline with plain lines in figure 2 the food yield gap that is absorbable in each world region by closing region-specific crop yield gaps and by suppressing feed-grain. The direction of plain lines indicates the relative change in N self-sufficiency for closing crop yield gaps. Regions with lower $r_{BNF}$ potential than in the productivity frontier (table 2), could close food yield gaps with lower N self-sufficiency than illustrated in figure 2. A calibration of N self-sufficiency per world region would necessitate knowledge on regional BNF gaps.

World regions are positioned twice in figure 2 because regional productivity embeds feed trade. To show the actual contribution of each world region to global food availability, the apparent food yield as calculated from national production data (circles) is corrected for feed trade (cross marks) by subtracting net feed imports from domestic primary production and by accounting for net feed exports in terms of food-equivalent. The food value of feed is a fraction proportional to the livestock conversion efficiency. N self-sufficiency increases with the correction for feed trade in net importing countries (see Supplementary section 6).

Net feed imports greatly vary among regions and are maximum for Western Europe (mainly due to soybean imports) and Eastern Asia (soybean and maize). In both regions, net feed imports account for about half total concentrate feed use. The actual food yield of these two regions is about half the apparent yield. Accordingly, Western Europe has the highest food yield potential worldwide but also the largest food yield gap. In Southern Europe and Northern Africa, the food yield after correction for feed trade is almost zero, meaning that those regions contribute almost no food to the world because their livestock sectors consume all net primary production. North and South America, which are the biggest agricultural net exporters worldwide, mainly export feed. We estimate that feed makes up at least 60% of total net exports, which reduces by half the contribution of North and South America in world food availability.

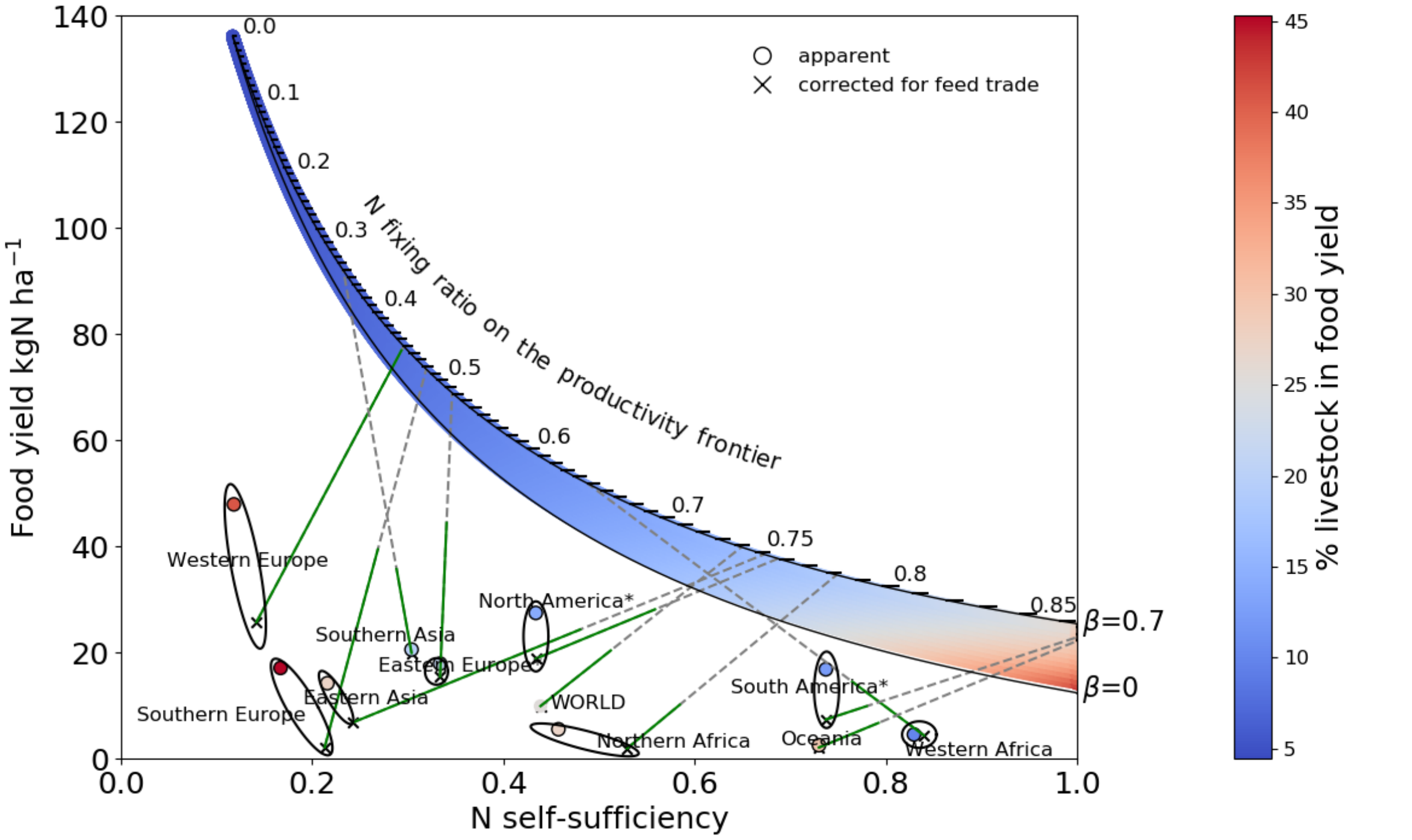

**Figure 2| The productivity frontier of agriculture** i.e. the maximum food yield, as a function of N self-sufficiency over the full range of the N fixing ratio (N fixing to total agricultural land) and of β (manure N recycling rate). The N fixing ratio puts an upper limit to food yield and a lower limit to livestock production (color code). World regions are projected on the productivity frontier for β = 0.7 (see country list per region in the Supplementary Table S4). World regions are shown both as per apparent food yield (circles) and after correction for feed trade (cross marks). Net feed exporting regions are marked with an asterisk. The plain green lines highlight region-specific food yield gaps according to region-specific crop yield gaps (see Supplementary section 6). Food yield gaps can be reduced by closing crop yield gaps and by reducing feed-grain.

## Defining world food production boundaries

Integrating yield gap analysis [30] with N use efficiency in world agricultural systems [26] lays the foundations for linking the N planetary boundary [15] with food production boundaries. Food production boundaries are ultimate limits to supportable human population and interface production with N loss and N self-sufficiency. We define five world food production boundaries (B1, B2, B3, B4a and B4b) as a function of feed-grain and N use efficiency in crop and livestock production. Three boundaries are $N_{ind}$-based considering a full closure of crop yield gaps worldwide (B1, B2, B3). Two boundaries are organic (B4) considering current BNF rate (B4a) and improved BNF rate (B4b). All boundaries consider total agricultural land equal as today, dismissing as unsustainable a net expansion of agriculture on forestland (See Methods and Supplementary Section 7). In each case, the model is parametrized by setting the variables to limit values considered achievable on the global scale (Table 2). Note that the global crop mix affects the maximum achievable crop yield. Shifts toward higher cereals production (which are the most productive crops) would tend to increase the maximum yield; in contrast, shifts toward more vegetables, fruits and nuts would tend to lower it. Current diets are relatively poor in vegetables, fruits and nuts, meaning that a transition toward healthier diets [28] equates to a global crop mix with lower yield potential than the current crop mix.

**Table 2| Driving variables and endpoint values used in the five world food production boundaries.** The values of $N_{atm}$, $NHI_{fixing}$ $NUE_{fixing}$, $HI_{majorcrops}$, feed-residues % are taken from table 1. All boundaries include potential land trade-offs (ΔS) of 100 Mha in favor of major crops land in $N_{ind}$-based boundaries (B1, B2, B3) and in favor of N fixing land in organic boundaries (B4a, B4b). The maximum major crops yield in $N_{ind}$-based boundaries is calculated based on *Mueller et al.* [30]. See Methods and Supplementary section 4 for NCE, section 8 and Figures S3 and S4 for $NUE_{crop}$ and major crops yield and, Section 9 and Fig. S5 for crop yield in organic boundaries.

| Boundaries | | B1 Maximum yield & current efficiency | B2 Maximum yield & crop efficiency | B3 Maximum yield & total efficiency | B4 Organic food & total efficiency | |
|---|---|---|---|---|---|---|
| Variables | Unit | | | | B4a | B4b |
| N fixing ratio | % | [74, 74 + ΔS] | [74, 74 + ΔS] | [74, 74 + ΔS] | [74, 74 + ΔS] | |
| $r_{BNF}^{fixing}$ | $kgN \cdot ha^{-1}$ | 15 | 15 | 15 | 15 | 25 |
| $r_{BNF}^{majorcrops}$ | $kgN \cdot ha^{-1}$ | 7.5 | 7.5 | 7.5 | 7.5 | 7.5 |
| $N_{ind}$ | $kgN \cdot ha^{-1}$ | variable | variable | Variable | 0 | |
| $Yield_{majorcrops}$ | $kgN \cdot ha^{-1}$ | 70 | 70 | 70 | variable | |
| $NUE_{crop}$ | % | 38.5 | 70 | 70 | 70 | |
| Feed-grain | % | [0, 70] | [0, 70] | [0, 70] | [0, 70] | |
| Livestock NCE | % | 12.5 | 12.5 | 12.5 | 12.5 | |
| Manure recycling rate | % | 24 | 24 | 70 | 70 | |

## *Framing the world trajectory*

Figure 3a brings together world food production boundaries, human population projections and the historical world trajectory of food production, N loss and self-sufficiency. The N planetary boundary indicates the range of allowable N loss from agriculture (high, medium and low) required to remain within the safe N operating space (see Supplementary section 10). The historical world trajectory is calculated from FAOSTAT data and is successfully reproduced by the model for the current N regime (averaging 2010-2013 data). Since 1961, the N food yield has almost tripled while N self-sufficiency dropped twofold from 0.8 to 0.42. The fact that the average N food supply per capita increased since 1961 is highlighted by the irregular graduation in the axis of human population in the inset figure in Fig3a. The model demonstrates that in 2013, $N_{ind}$ supported 65 % of the world population. This is consistent with the estimate of 40% in the mid-1990s by *Smil* [5] given the subsequent 30% population growth, and significantly higher than the 44% reported in *Erisman et al*. [4] for 2008. The world food trajectory crossed the high end of the N planetary boundary in the 1970s.

Food yield, N loss and $NUE_{tot}$ relate through this equation: Food yield = $N_{loss}$ x $NUE_{tot}/(1-NUE_{tot})$ where the term $NUE_{tot}/(1-NUE_{tot})$ is the slope of the world trajectory in Figure 3a. With current $NUE_{tot}$, N loss outstrips the food yield four-fold. The global trend is an aggregate of parallel improvements in the $NUE_{tot}$ in some countries[31,32] and unsustainable intensification in other countries[26]. Future trends will largely depend on whether sustainable intensification will be achieved at early development stages in developing countries, in particular China and India, which together account for 43% of global $N_{ind}$ use. For illustration, the joint $NUE_{crop}$ increase in China and India from currently 30% [33] to 40% would lift-up global $NUE_{crop}$ by 5%. The goal of the Chinese central government is to reach zero increase in the use of fertilizers and pesticides by 2020 [34]. However, it is unclear whether Chinese agriculture is close to turning the corner of environmental N efficiency to fulfill that goal.

## *Maximum supportable human population*

The five boundaries explore the biophysical limits of food production regardless of potential socio-economic constraints. Depending on the boundaries and the share of feed-grain, we show that the maximum supportable human population varies from 6 to almost 17 billion people (Fig. 3a, right axis). The population range is an endogenous variable in the model and puts population forecasts as derived from natality-mortality dynamics in perspective with food resources. Only a handful of previous analysis have challenged human demographic projections in light of resource availability constraints [35–37].

We show that the projected human population in the 21$^{st}$ century (2050 and 2100 low, medium and high variants [38]) can only be conditionally sustained independently of any timeframe (Fig. 3a). This means that regardless of potential growth rates in the variables underpinning food production, there is high risk of food shortages unless there is a net expansion in agricultural area or a reduction in individual demand. World average food yield per capita is almost twice the recommended N intake, so reducing loss is a major lever for improving food security. However, analysis of food consumption patterns suggests that food loss increases with income and, therefore, increasing economic welfare in currently low-income countries is likely to exacerbate N loss (See Methods and Supplementary Section 11 and Fig. S6). Our calculations consider the current global average share of loss in food yield. A reduction in loss could require a global-scale humanitarian awareness on food issues and security.

We show that, across all $N_{ind}$-based boundaries (B1, B2 and B3), feed-grain competes with human nutrition such that the maximum supportable human population more than doubles when feed-grain drops from 70% to 0 (Fig. 3a). Feed-grain of 70% is currently at stake in Western Europe and 46% is the world average. Since the 70s, the share of feed-grain in global grain production has reduced from 56 to 46 % but, in terms of mass, feed-grain has increased threefold, driven by the increase in crop yields and major crops land (Supplementary section 12 and Fig. S7). For the current global share of feed-grain, the $N_{ind}$-based boundaries (B1, B2, B3) would support 9.8 billion people, which is the medium 2050 population variant. If the feed-grain share was to increase, this level could not be fulfilled. The low 2050 variant of about 8.7 billion people is compatible with a feed-grain share of up to 56%. The high 2100 population variant is out-of-reach even in the case of a full suppression of feed-grain, and would require, in addition, a land trade-off of 100 Mha in favor of major crops land. Feed-grain suppression is unrealistic with respect to current diets as it would equate to 15% instead of 35% of animal proteins in food supply (see Supplementary section 13 and Fig. S8). It is even inadvisable with respect to healthy diets, which recommend 25% of animal proteins in food supply [28]. The 2050 high and the 2100 medium population variants would require feed-grain to decrease to about 30%. This would support average animal proteins production close to the recommended value of 25%, thus allowing to reconcile a health objective with a highly probable end-of-century demographic figure.

### *Food yield gaps in organic boundaries*

There is no consensus in current literature about the feeding capacity of organic farming [39,40]. N limitation is a matter of $r_{BNF}^{fixing}$, feed-grain and $NUE_{tot}$. Unlike in $N_{ind}$-based systems, food yield in organic systems is directly governed by $NUE_{tot}$ because N loss is hardly compensated by additional N input. We find that the crop yield is 41 kgN·ha-1 in B4a (current $r_{BNF}^{fixing}$) and 58 kg N·ha$^{-1}$ in B4b (improved $r_{BNF}^{fixing}$), implying that crop yield gaps between organic (B4a and B4b) and $N_{ind}$-based boundaries (B1, B2, B3) are 42 and 18 % respectively. Both values are consistent with the scarce on-field data on organic and conventional crop yields under various production conditions[41]. Lower yields in the organic boundaries explain why the population range is narrower than in $N_{ind}$-based boundaries. As demonstrated by B4a, an increase in the world $NUE_{tot}$ from 24 to 50 % (fig 3a and 3b), could entirely offset $N_{ind}$ input to agriculture and still sustain 6.7 billion people with the current share of feed-grain. In B4b, the combination of improved $NUE_{tot}$ and $r_{BNF}^{fixing}$ could respond to the medium 2050 population variant of 9.8 billion people with the current share of feed-grain. Inversely, the organic version (no $N_{ind}$) of world agriculture with current feed-grain and current $r_{BNF}^{fixing}$ and $NUE_{tot}$ could hardly feed 3 billion people (Figure 3a).

### *Effect of feed-grain in $NUE_{tot}$*

The slope in each boundary is the ratio of food yield to N loss and highlights the effect of feed-grain in $NUE_{tot}$. Naturally, $NUE_{tot}$ decreases with feed-grain, but the lower the absolute level of N loss, the greater the relative decrease in $NUE_{tot}$ (Fig 3b). This explains why the decrease in $NUE_{tot}$ is the sharpest in B3, the mildest in B1 and in-between in B2. As a consequence, the reduction in feed-grain in countries with relatively high $NUE_{crop}$ would result in a higher increase in $NUE_{tot}$ than in countries with low $NUE_{crop}$ (see Supplementary Fig. S9 for $NUE_{tot}$ as a function of $NUE_{crop}$, N manure recycling and feed-grain). The comparison between $NUE_{tot}$ in the boundaries and the world trajectory (figure 3b) reveals that the maximum potential level of $NUE_{tot}$ is twice as high as today.

B4a is the only boundary to entirely fit within the safe N operating space for humanity. B3 and B4b are below the average value of the N planetary boundary and approach the lower end when feed-grain is zero. B2 falls within the N planetary boundary for feed-grain below 40 %. In B1, N loss exceeds by two to three times the allowable average N loss depending on feed-grain. Note that $NUE_{tot}$ greatly influences the requirement in $N_{ind}$. With the current feed-grain, $N_{ind}$ is about 60 % higher than today in B1 and 60 % lower than today in B3 (see Supplementary Fig. S10). The interspace between the five boundaries and the current N regime designates intermediate situations of total N input and $NUE_{tot}$ in the N operating space.

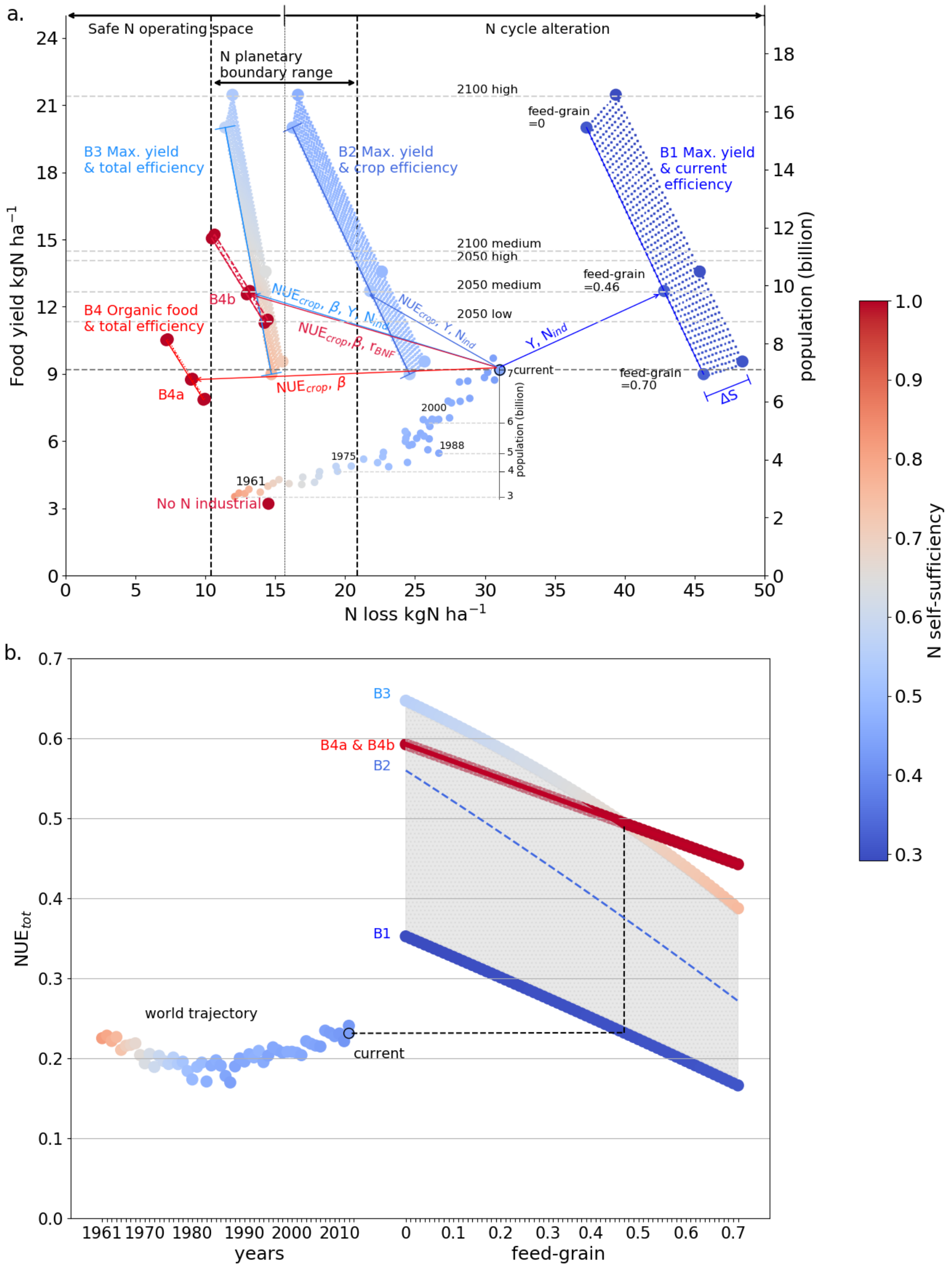

**Figure 3a| Food yield, N loss and N self-sufficiency (color code) for world agriculture as a function of feed-grain in the five boundaries**. The world trajectory since 1961 and the hypothetical case of current agriculture without industrial N fertilizers are given for comparison. The historical trajectory takes into account the increase in N food supply per capita over time, leading to the irregular graduation in the population axis (inset figure in 3a). All boundaries include potential land trade-offs (ΔS) of 100 Mha in favor of major crops land in $N_{ind}$-based boundaries (B1, B2, B3) and in favor of N fixing land in organic boundaries (B4a, B4b). Food yield is linked to supportable world population by considering current food supply per capita. The food production boundaries put in perspective supportable population with 2050 and 2100 population projections. The driving variables that link the current regime to each boundary are indicated on the connecting arrows. The N planetary boundary corresponds to the range of allowable N loss from agriculture to stay within the safe N operating space according to *Bodirsky et al.* [10] and *De Vries et al.* [11]. **Figure 3b| $NUE_{tot}$ in the world trajectory since 1961 and $NUE_{tot}$ in the boundaries as a function of feed-grain.**

## Growth rate challenges for achieving goals on time

The five boundaries provide endpoints in food yield and N loss and are not intended as speculation on trajectories. Yet, the growth rates in production and efficiency become critical for tackling time-specific goals. Potentially disastrous tipping points for the biosphere may lurk behind the prolonged alteration of the global N cycle, but because tipping points depend on so far largely unknown complex system feedbacks, no time frames for N sustainability are available in the literature. In contrast, the time-interlinkage between food yield and demand is easier to perceive and presents as the Malthusian trap. Given that the medium 2050 UN population variant cuts across the food production boundaries (B1, B2, B3, B4b) at the current share of feed-grain, hitting this target necessarily echoes policy agendas. Figure 4 provides the growth rates required for production and efficiency at the global scale (see Supplementary Table S5) to arrive at B3 and B4b by 2050. The calculated growth rates are put in perspective with records from selected countries, the world average and China as the biggest crop producer worldwide over the last decades[20].

Not surprisingly, the most critical challenge will be in the synchronous increase of major crops yield and $NUE_{crop}$, which only five countries have managed to achieve over the last 30 years (Fig 4 and supplementary Fig. S4). Considered alone, average major-crops yields have grown in recent periods (2002-2011) even faster than required to meet the 2050 food demand [42] from this point forward. However, during past increases, the $NUE_{crop}$ typically decreased [33]. Similarly, the growth rate required of global NCE is of the same magnitude as that observed since the 1950s [22], but henceforth, it must build on livestock systems coupled with crop systems to limit manure N loss. The opposite trend, i.e. massive manure N loss from livestock decoupled from crops is an environmental side effect of the NCE increase within a context of economies of scale in 20$^{th}$ century agriculture.

The growth rates in figure 4 are overwhelmingly from $N_{ind}$-based systems. The comparison with organic systems should be handled with caution. Currently, organic agriculture is highly dynamic with the growth rate in organic land approaching 20% between 2016 and 2017. However, the current organic acreage of roughly 70 Mha worldwide [43] is tiny compared to the 4.7 billion hectares of total agricultural land and it is unclear whether a scale-up of this magnitude is feasible, and at what rate and cost. Even if it were, agricultural farming today involves practices that do not match the specifications in B4b meaning that no extrapolation toward B4b can draw on current organic land growth or global historic trends. Firstly, a considerable portion of total N input to organic systems today comes from conventional systems[44], which violates the constraint of N self-sufficiency. Secondly, both the yield and $NUE_{tot}$ are below the requirements in B4b [41,45]. Thirdly, current world $r_{BNF}^{fixing}$ lags behind the required value in B4b even though remarkable growth rates has been observed by the past [12,32]. Notwithstanding, the fact that organic farming is structurally limited by N makes food security and N loss abatement goals synergistic, forming a common incentive for improving $NUE_{tot}$.

Overall, food goals seem to be more tangible than N sustainability goals and, unfortunately, the world trajectory seems to be on track toward B1 (fig 3a). This is coherent with the fact that with current feed-grain, achieving B1 by 2050 would require a growth rate in $N_{ind}$ use equal to the rate observed since 2000 (see Supplementary Fig. S11). Of course, the challenge is uneven among countries and food security around the world will largely depend on international trade. For instance, in several world regions, food demand is expected to exceed food production even with a full closure of crop yield gaps. Sub-Saharan Africa is probably the most critical case with a projected population growth by a factor of 2.5 by 2050 [38] against a crop yield gap of 2.2 [30]. Conciliating the food production boundary and the N planetary boundary, would take a global-scale turnaround to increase average food yield by 50 % while achieving world average $NUE_{crop}$ and manure N recycling of 70%, equating to a $NUE_{tot}$ of roughly 50% against 24% today. This would necessitate highly restrictive agro-environmental policies in support of a rapid implementation of sustainable crop and livestock practices in all currently low-efficiency agricultural systems worldwide. Actions in favor of improved $NUE_{tot}$ will be increasingly pressing as human population increases, since N loss is always a correlative of the amount of N in the food yield.

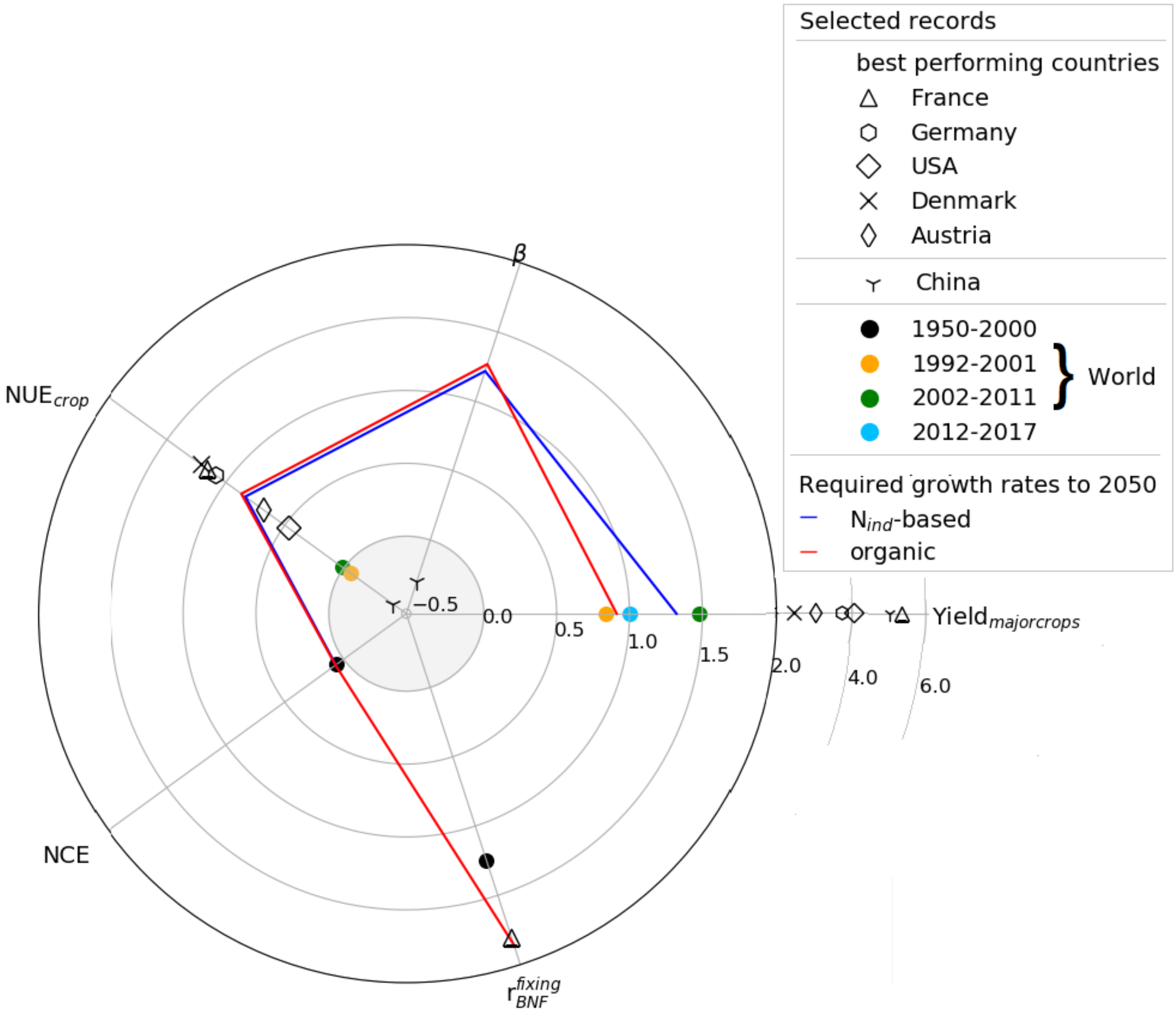

**Figure 4| Average annual growth rates (% yr$^{-1}$) required in the key agricultural variables governing production and efficiency to achieve the food and N planetary boundaries by 2050 with the current share of feed-grain (B3 and B4b with baseline 2013).** The required growth rates are compared to selected records at national and global scales over the last decades. National-scale rates are calculated for 1961-2017 for $NUE_{crop}$, 1980-2000 for manure N recycling (β) and in the aftermath of World War II for $r_{BNF}^{fixing}$. The grey zone indicates negative rates. All rates are non-compounding and are calculated in supplementary Table S5 and Fig. S4.

## Methods

### *Agricultural land*

Total agricultural land is calculated from 1961 to 2013 as the sum of all individual land grazed or harvested reported in FAO plus the land cultivated with fodder legumes (Supplementary S1). Total agricultural land is distinguished between N fixing land and major crops land. N fixing land includes grasslands, fodder legumes, soybeans, and groundnuts while major crops land includes all other crops grown for food, feed or other uses. The N fixing ratio i.e. the ratio of N fixing land to total agricultural land (see Supplementary section 1 and Fig. S1) amounts to 74% at the global scale today after a slight decrease since 1961 and is projected to remain unchanged to 2030 [19]. It largely controls the food yield and determines most of the potential BNF transfer to major crops land through crop rotations and manure recycling.

### *BNF calculation*

Based on BNF data from literature aggregated per land category, we estimated BNF inputs to agriculture from N fixing land and major crops land at regional and global scales. For soybean, groundnut and all pulses, BNF rates are estimated as a function of crop yields using the methodology in *Lassaletta et al.* [33] and FAOSTAT data on crop yields and harvested areas [20]. For sugar cane and rice, BNF rates are considered constant at respectively 25 and 33 kgN·ha$^{-1}$ [46]. For grasslands we used constant BNF rate of 5 kgN·ha$^{-1}$. BNF in grasslands is particularly uncertain due to a large diversity of land types, including intensively managed and unmanaged pastures, savannahs and rangelands [47,48]. A recent analysis[49] estimated average BNF rate in permanent grassland in Europe at 6.8 kgN·ha$^{-1}$. For fodder legumes, BNF estimates at the global scale[46] as reconstructed from *Smil* [7] are distributed at the scale of world regions by weighting with cattle stocks distribution data [20]. In total, global agricultural BNF is estimated at 61 TgN in 2013 and falls within the 50-70 TgN range in *Herridge et al.* [46]. N fixing land accounts for 85% of total BNF (52 TgN) with average BNF rate ($r_{BNF}^{fixing}$) of 15 kgN·ha$^{-1}$. The remaining 15 % (9 TgN) occurs on major crops land under cultivation of pulses, rice, and sugarcane with average BNF rate ($r_{BNF}^{majorcrops}$) of 7.5 kgN·ha$^{-1}$. The distribution of total agricultural BNF across world regions per land category and according to biomass allocation between food and feed is summarized in Supplementary Table S1, Section 2 and Table S2.

### *N budget model*

The model accounts for the main N flows in agricultural systems connecting N fixing land, major crops land and livestock production systems through N balances in each compartment. The set of N balance equations is given in the Supplementary Section 3. Total N input ($N_{tot}$) to the agricultural system is the sum of $\mathrm{BNF}, \mathrm{N_{atm}}, \mathrm{N_{ind}}$ and $\mathrm{N_Feed_{net-imp}}$ where BNF is the sum of BNF in N fixing land and major crops land, $N_{atm}$ is the atmospheric N deposition, $N_{ind}$ is industrial N fertilizers and $N_Feed_{net-imp}$ is the net N feed imports (relevant at regional scales). $N_{tot}$ divides into food yield and loss according to total N use efficiency, assuming no change in the soil N pool and no N recovery from human excreta or retail food waste. Food yield is total human edible production per agricultural hectare.

### *Food yield and total N use efficiency*

Food yield is the amount of N extracted in primary production plus livestock production minus feed (including crop residues). Food yield accounts for the overwhelming majority of all N extracted because fiber constitutes only 2.7% of total N in world primary crop production [20] while biofuels are negligible. Food yield is also largely controlled by crop yield gaps [30] and by allocation of grain production to feed. Overall, the food yield is the consequence of $N_{tot}$ to agriculture and total system N use efficiency ($NUE_{tot}$). $NUE_{tot}$ is the result of field NUE and of the N recycling efficiency in livestock systems. While improving field NUE requires the implementation of technologies and management practices at the farm scale [26], improving $NUE_{tot}$ additionally involves dietary choices and efficient nutrient recycling from livestock to crop farms. $NUE_{tot}$ has been given little scrutiny in the literature. High $NUE_{tot}$ benefits to N self-sufficiency and is essential to the food yield of systems that restrict or ban the use of $N_{ind}$, such as organic agriculture.

### *Crop nitrogen use efficiency*

Crop nitrogen use efficiency ($NUE_{crop}$) and the way it relates to total N input are well documented for major crops land. Under steady production practices, $NUE_{crop}$ is a decreasing function of fertilization [33], whereas under improving production practices, crop yields can increase even with decreasing fertilization; there are numerous examples of countries that have turned the corner of an environmental Kuznets curve in terms of $NUE_{crop}$ [26]. Although the upper limit of $NUE_{crop}$ is unclear, the historical trajectories of countries combining both high yields and efficiency indicate that $NUE_{crop}$ tends to an asymptote at around 70 to 80 % [33] and Supplementary S8). Maximum observed crop yields approach 130 kgN·ha$^{-1}$ which is also the maximum value obtained from crop yield gap analysis [30]. On average, for the current global crop mix, N yield and $NUE_{crop}$ in world cropland are estimated at respectively 56 kgN·ha$^{-1}$ and 42 % [26]. When compiled for the major crops land defined in this Article, the world average N yield and $NUE_{crop}$ are 47 kgN·ha$^{-1}$ and 38 % respectively.

### *Livestock nitrogen use efficiency*

N use efficiency in livestock production depends on the N conversion efficiency of feed to food (NCE) and of the rate of manure N recycling into crop production. Manure N recycling reflects crop and livestock integration in agricultural systems. Based on *Smil* [7], we estimate global average manure N recycling at 24% which closely aligns with a recent FAO publication [50] while falling within the range indicated by *Sheldrick et al.* [51] and *Oenema and Tamminga* [52]. Some argue that manure N recycling could ultimately increase to 80% [34]. NCE greatly varies among livestock production systems from about 5 % in beef to 10-15 % in pig, 20-30 % in chicken and 30-40 % in milk and eggs production [23]. Based on *Bouwman et al.* [22] and *Billen et al.* [21], average world NCE is estimated at 10%. Global average NCE can increase through improved livestock rations as well as through dietary changes toward livestock products with higher NCE.

### *Feed calculation from FAOSTAT data*

FAOSTAT data (1961-2013, [20]) allow estimating N feed at regional and global scales. Note that feed in FAO commodity balance sheet does not include crop residues, grasslands and fodder legumes. Calculation approaches differ in the literature. One approach consists in calculating N feed as a function of livestock production and livestock conversion efficiency combined with assumptions on feed from crops residues and grasslands [10,15,16]. Another approach consists in deriving N feed from biomass balances in FAO Food Balance Sheet [13,53]. In this latter approach, assumptions are needed to allocate N among vegetal food supply, seed, retail food loss and non-food uses (such as soap, pet food). In *Lassaletta et al.* [13], N feed is the difference between available crops production (crop production plus net imports minus interannual change in stock (Δstock)) and final human consumption and thereby, feed implicitly includes N in seed, other uses and retail food loss. The calculation uses a set of N contents for crops production [54,55] and a set of N contents for food supply [56]. Our approach is similar to *Lassaletta et al.* [13] with the difference that seed, other uses and retail food loss are not accounted for as feed. They are subtracted from the feed calculation by assigning to seed and other uses the N content of crop production and to retail food loss the N content of vegetal food supply across all FAOSTAT categories *i* (eq 1).

$$N_feed = \sum_i \big(available\ production - (vegetal\ food\ supply + seed + processed + non-food\ uses + retail\ food\ loss)\big) \quad \text{(eq. 1)}$$

We consider as *Lassaletta et al.* [13] that all feed reported in FAO commodity balances is for terrestrial livestock production, therefore excluding fish farming.

We obtain that the share of feed in global N crop production is 56 % in 2013 which is in agreement with the detailed analysis in *Cassidy et al.* [53], reporting global feed share of 53 % in terms of proteins, 24 % in terms of mass and 36% in terms of calories. From all grain production, the share of grain allocated to feed (feed-grain) is estimated at 48 % at the global scale against 66 % in *Lassaletta et al.* [13]. Feed-grain per world region is summarized in Supplementary Table S2 (column 9).

### *Maximum food yield per agricultural hectare*

By setting the model variables that govern production and efficiency to the most optimistic values gathered in Table 1, we define the productivity frontier of agriculture as the theoretical maximum food yield across the N fixing ratio. On the one hand, the selected values maximize land productivity while minimizing resource use beyond what is at best observed worldwide. On the other hand, feed-grain is set at zero, which eliminates feed competition with food and in practice would limit any production from monogastric animals to recovery of feed from waste.

*N food production, supply and intake*

The maximum supportable world population is calculated for the current average N food production per capita which has increased over time. The recommended N intake is currently two thirds of the average N food supply as reported by FAO (see Supplementary Fig. S6a for the distribution of N food supply over total population) and about half the N food production (see Supplementary section 11). This difference is due to food loss, including post-harvest loss [57], overconsumption and non-food uses. The potential increase in global food availability by decreasing food loss is substantial. However, it has been shown that food loss increases with income [58] and supplementary Fig. S6b for a classification by income of N food loss) and therefore increasing economic welfare in currently low-income regions might exacerbate global N loss.

*Crop yield gaps closure in $N_{ind}$-based boundaries*

$N_{ind}$-based boundaries (B1, B2, B3) assume full closure of crop yield gaps in all world regions [30]. The requirement of $N_{ind}$ is a function of crop yields, feed-grain and $NUE_{tot}$. Considering the current distribution of cropland, crop yield gaps and crop mix worldwide, we obtain a weighted maximum major crops yield of 70 $kgN \cdot ha^{-1}$ (Table 2). While it could be argued that improved crop cultivars could break this biophysical barrier, climate change is likely to have a negative impact on crop yields [59,60]. In addition, a shift toward healthier diets would lower the yield potential of major crops land because it would require less cereals and more vegetables, fruits and nuts with lower yields potential than cereals.

*Yields in organic boundaries*

The two organic boundaries (B4a and B4b) differ in the value of $r_{BNF}^{fixing}$ which drives total N input to agriculture. Although there is no clear evidence about intrinsic yield gaps between organic and conventional cropping systems for similar levels of N input [25,41,45], N is the main limiting factor in organic systems and, therefore, organic yield gaps may be more challenging to close [61]. Currently, N fixing land contributes 52 TgN to agriculture, which is the value used in B4a. Although there are no global estimates of the BNF potential in agriculture, one of the levers to increase BNF is to replace current fallow land and parts of unmanaged grasslands with grain and fodder legumes. The 19$^{th}$-century agricultural revolution in Western Europe built on this practice [62]. At the global scale, fallow land is primarily located in developing countries in Southeast Asia and Africa, and is estimated between 86 and 200 Mha [20,63] . The plantation of all current fallow land with highly fixing fodder legumes (120 $kgN \cdot ha^{-1}$) would supply additional 14-18 TgN to agriculture. Additionally, by converting 5 to 7 % of the world’s unmanaged grasslands to fodder legumes, BNF would further increase by 37 TgN, meaning that total BNF from N fixing land would be 89 TgN. This is the value used in B4b, corresponding to global average $r_{BNF}^{fixing}$ of 25 $kgN \cdot ha^{-1}$.

*Gaps in nitrogen use efficiency*

The improved values of $NUE_{crop}$ and manure N recycling in boundaries B2, B3, B4a and B4b can be considered very ambitious for global agriculture. On the one hand, only a handful of countries (the United States of America, Austria, Denmark, Germany and France) have managed to achieve both crop yields above 70 $kgN \cdot ha^{-1}$ and $NUE_{crop}$ above 65% (see Supplementary section 8 and Fig. S3, Fig. S4 for the current situation and historical trends). On the other hand, improved manure N recycling would require a much tighter spatial integration between crop and livestock systems, which goes against current trends. The share of recycled manure has most likely decreased inversely to agricultural specialization and economies of scale. Economies of scale have driven, for the last half-century, the decoupling of crop and livestock systems [64] and even the emergence of totally footloose livestock operations entirely dependent on purchased feed [65]. The emergence of economies of scale in the ongoing agricultural transition from traditional to industrial systems in developing countries is very likely to further reduce global manure N recycling. A decrease from 33 to 21 % over three decades has been recently estimated for China[66].

*Agricultural land trade-offs in the boundaries*

In all boundaries total agricultural land is considered equal to today. This may be seen as an optimistic assumption, especially if forestland were to be preserved for environmental sustainability. Today, there are serious concerns about land loss due to urban expansion [67], climate change forcing, soil degradation [63] or a combination of these factors. Over the past decades, total agricultural land has slightly increased due to deforestation especially in the tropics [68] and has tended to stabilize in recent years (supplementary Fig. S1). Based on integrative planetary system analysis for key ecosystem services, we dismiss as unsustainable the case of a net expansion of agriculture on forestland [69,70] (see Supplementary section 7). Within this constant agricultural land, we consider land trade-offs that increase productivity. Starting with the current global N fixing ratio, we limit land trade-offs to 100 Mha, a 8 % change already discussed as plausible [71], in favor of N fixing land in the organic boundaries (B4a, B4b) and in favor of major crops land in $N_{ind}$-based boundaries (B1, B2, B3). In the organic boundaries where total N input is governed by BNF, crop yields are a function of the N fixing ratio (Supplementary section 9 and Fig. S5).

Data availability: The main input data used in the analysis are taken from the statistics of the United Nations Food and Agriculture Organization (FAO), available at http://www.fao.org/faostat/en/#data. Nitrogen content coefficients are taken from reference [55]. Crop yield gap data are taken from reference [30]. Other literature sources to calibrate model parameters are specified in Table 1 of the manuscript. The underlying data and results regarding the positioning of world regions in the productivity frontier in 2013 are given in Supplementary Table S2.

Supplementary Materials: The following are attached below. The supplementary information provides details on the model structure and on data sources and methods used in the calculations.

Conflicts of Interest: The authors declare no conflict of interest.

Funding: The work built on funds of the research program Emergence Ville de Paris Convention 2015 DDEES 165.

## Supplementary Information

This supplementary information provides details on the model structure and on data sources and methods used in the calculations.

### 1. Agricultural land and the N fixing ratio

Total agricultural land is calculated from 1961 to 2013 as the sum of all individual land grazed or harvested reported in FAO plus the land cultivated with fodder legumes which is not reported in FAO. Land cultivated with fodder legumes is taken from *Herridge et al.* [1]. Accordingly, total agricultural land in our calculations is the sum of all productive land excluding fallow land. It sums at 4787 Mha in 2013, which is 95 Mha below the total agricultural land reported in the FAO land module[2]. The difference between the FAO land accounting and our calculations reduces in time following the progressive reduction in fallow land (figure S1).

Total agricultural land is distinguished between N fixing land and major crops land. N fixing land is the sum of soybean, groundnut, fodder legumes and grasslands while major crops land includes all other crops grown for food, feed or other uses. The ratio of N fixing land to total agricultural land is the N fixing ratio ($NFixing_{ratio}$). While total agricultural land has increased since 1961 by 10% [2] and productive land excluding fallow land by 16%, the N fixing ratio has remained stable in time (Figure S1). Note that the rate of growth in agricultural land slowed down from about 0.36% $yr^{-1}$ between 1961 and 2000 to about 0.1% $yr^{-1}$ since 2000.

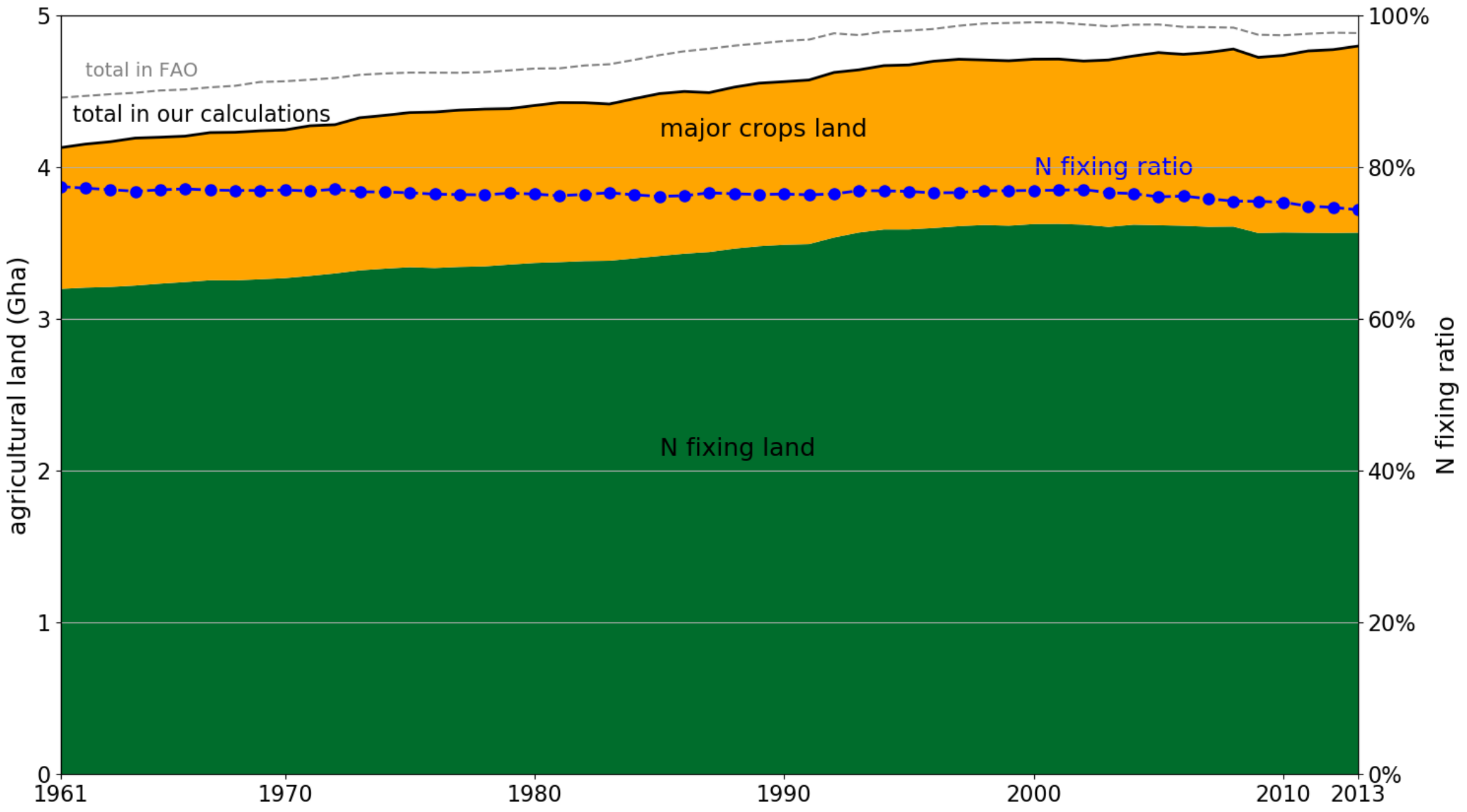

**Figure S1| Total agricultural land (left axis) and the N fixing ratio (right axis) in time at the global scale. Total agricultural land is distinguished between N fixing land and major crops land.** The N fixing ratio is the ratio of N fixing land to total agricultural land. Total agricultural land in FAO includes temporary fallow land and excludes the land cultivated with fodder legumes, which explains the difference in total agricultural land between our calculation and the FAO land module.

## 2. BNF distribution among world regions

Table S1 summarizes the distribution of BNF across world regions per land category and according to biomass allocation between food and feed. The distribution of BNF per world region is supported by the comparison with specific regional and country level assessments based on national land use databases. For instance, our BNF estimate for North America (9.5 TgN) closely matches the sum of BNF estimated separately for Canada [3] and the United States of America [4]. BNF for Asia (20.4 TgN) is close to a previous estimate of 18.8 TgN [5]. BNF in South America in 2013 (16 TgN) is close to the sum estimated separately for Brazil [6] and Argentina [7] in 2005 and 2006, given that soybean area has increased by 55 % in both Brazil and Argentina since the studied dates and that these two countries account for 80% of agricultural land in South America.

Note that BNF is in decline in regions that heavily rely on feed proteins imports from other regions. For instance in Europe, grain legumes have lost 3.8 Mha (65%) since 1961 [8] following the abandonment of domestic grain and fodder legumes cultivation. BNF per world region can be derived from Table S2 as $BNF = agricultural\ land \times \left[NFixing_{ratio} \times r_{BNF}^{fixing} + (1 - NFixing_{ratio}) \times r_{BNF}^{majorcrops}\right]$ (columns 3 to 6).

**Table S1| Global BNF distribution between N fixing land and major crops land (based on year 2013) and between food and feed.** Sources: [(1)] Calculated based on *Herridge et al.* [1] and *Lassaletta et al.* [9] [(2)] FAO data [2] except for fodder legumes [1]. Grasslands corresponds to the FAO permanent meadows and pasture category.

| Products category | Land type | Biomass use | BNF rate[(1)] kgN ha$^{-1}$ | Land[(2)] M ha | BNF TgN | % of total BNF |
|---|---|---|---|---|---|---|
| Soybean and groundnut | N fixing land | LV feed | 158 | 138 | 22 | 36 |
| Fodder legumes | N fixing land | LV feed | 120 | 110 | 13 | 21 |
| Grasslands | N fixing land | LV feed | 5 | 3,309 | 17 | 28 |
| Pulses | major crops land | Human food | 38 | 84 | 3 | 5 |
| Rice paddy and sugar cane | major crops land | Human food | 32 | 191 | 6 | 10 |
| Cereals (excluding rice) | major crops land | Human food / LV feed | non-fixing | 558 | - | - |
| Other oilseeds | major crops land | Human food / LV feed | non-fixing | 143 | - | - |
| Roots and Tubers | major crops land | Human food / LV feed | non-fixing | 61 | - | - |
| Vegetable, Treenuts, Fruits, Others | major crops land | Human food | non-fixing | 192 | - | - |
| | | N fixing land (LV feed) | 15 | 3,557 | 52 | 85 |
| | | major crops land (human food and LV feed) | 7.5 | 1,230 | 9 | 15 |
| | | Total agricultural land | | 4,787 | 61 | 100 |
| | | % Fixing to total agricultural land (N fixing ratio) | | 74.3 | | |

## 3. Model structure

The model accounts for N flows in agricultural systems connecting N fixing land, major crops land and livestock production systems through N balances in each compartment. This section provides the set of N balance equations, on which the model is built, that link total N input ($N_{tot}$) to transformation processes towards products and loss from the agricultural system.

We define the food yield ($Food\ yield$) as total human edible food production per agricultural hectare. The food yield is modeled as the sum of major crops production ($Yield_{majorcrops}$) plus total livestock production (N_LV) minus feed from major crops land using equation 1.

$$Food\ yield = N_LV + Yield_{majorcrops} \times (1 - NFixing_{ratio}) \times (1 - \alpha_{grain}) \text{ (eq. 1)}$$

where $\alpha_{grain}$ is the share of major crops used as feed (feed-grain).

N_LV in equation 1 includes livestock production based on both N fixing land and major crops land and is calculated as the product of NCE and total feed ($N_Feed$) following equations 2 and 3:

$$N_Feed = Yield_{fixing\ land} \times NFixing_{ratio} + \left(\alpha_{grain} + \alpha_{residues} \times \frac{1-HI}{HI}\right) \times Yield_{majorcrops} \times (1 - NFixing_{ratio}) \text{ (eq. 2)}$$

$$N_LV = NCE \times N_Feed \text{ (eq. 3)}$$

where $Yield_{fixing\ land}$ is the yield in N fixing land, $\alpha_{residues}$ is the share of major crops residues used as feed and HI is the harvest index of major crops, i.e. the ratio of crop yield to crop's total aboveground biomass.

$Yield_{fixing\ land}$ is calculated using equation 4.

$$Yield_{fixing\ land} = N_{atm} \times NUE_{fixing} + r_{BNF}^{fixing} \times NHI_{fixing} \text{ (eq. 4)}$$

where $NUE_{fixing}$ is the N use efficiency, $NHI_{fixing}$ the N harvest index and $r_{BNF}^{fixing}$ the BNF fixing rate in N fixing land.

The share of BNF that is not recovered in feed is either transferred to major crops land through crop rotations ($BNF_{in\ rotation}$) or is lost in the environment ($N_loss_{fixing}$). The calculation follows the equations 5 and 6.

$$BNF_{in\ rotation} = NUE_{fixing} \times r_{BNF}^{fixing} \times (1 - NHI_{fixing}) \times NFixing_{ratio} \text{ (eq. 5)}$$

$$N_loss_{fixing} = \left(1 - NUE_{fixing}\right) \times \left(N_{atm} + r_{BNF}^{fixing} \times \left(1 - NHI_{fixing}\right)\right) \times NFixing_{ratio} \text{ (eq. 6)}$$

$Yield_{majorcrops}$ is the productivity per hectare of major crops land and is constrained by biophysical factors (see section 6). It equals total N input to major crops land ($Fertilization_{majorcrops}$) times the N use efficiency ($NUE_{crop}$). $Fertilization_{majorcrops}$ is the sum of $N_{ind}$, BNF in major crops land ($r_{BNF}^{majorcrops}$), manure N recycling from N fixing land to major crops land, BNF transfer from N fixing to major crops land through rotations ($BNF_{in\ rotations}$), and $N_{atm}$.

Manure N recycling from grain-fed livestock is addressed as a throughput within major crops land. Manure N recycling ($N_{manure_recycling}$) is a function of the recycling rate (β), the NCE in livestock production and total feed intake ($N_Feed$) (equation 7)

$$N_{manure_recycling} = \beta \times (1 - NCE) \times N_Feed \text{ (eq. 7)}$$

N loss from livestock ($N_loss_{LV}$) is the remainder of $N_Feed$ and is calculated from equation 8.

$$N_loss_{LV} = N_{manure_recycling} \times (1 - \beta)/\beta \text{ (eq. 8)}$$

N loss in major crops land ($N_{losses_{majorcrops}}$) is calculated from equation 9:

$$N_{losses_{majorcrops}} = \left(1 - NUE_{crop}\right) \times Fertilization_{majorcrops} \text{ (eq. 9)}$$

Total loss in agricultural land ($N_{loss}$) is the sum of N loss from each compartment calculated as follows: $N_{loss} = N_{loss_{LV}} + N_{loss_{majorcrops}} + N_{loss_{fixing}}$

We calculate total N use efficiency ($NUE_{tot}$) and N self-sufficiency of the agricultural system, as follows:

$$NUE_{tot} = \frac{Food\ yield}{N_{tot}} \text{ and } N_{self-sufficiency} = \frac{BNF + N_{atm}}{N_{tot}}$$

The model is implemented in Python and is validated on FAOSTAT data[2] for the year 2013.

## 4. Livestock conversion efficiency

Maximum nitrogen conversion efficiency (NCE) in livestock production is for dairy cows. Smil [10] provides ranges of NCE per livestock category from as low as 5 to 8 % for beef to as high as 30 to 40 % for milk and eggs. We consider NCE of 20% in the productivity frontier assuming 80% milk and 20% beef in livestock production for maximum beef and milk production efficiency.

The five food production boundaries are assessed by considering an increase by 25% in NCE from currently 10% to 12.5%. This increase is in line with commonly accepted potential gains through improved animal breeding and animal rations [11,12]. Nonetheless, this increase is possibly optimistic in the calculation of the boundaries where 35% of all crop residues are used as feed. Crop residues have a lower nutritional value than grain and, therefore, can typically support lower NCE [13]. The same conclusion is supported by historical analysis of animal rations and productivity [14,15]. Accordingly, livestock production in the boundaries is calculated with both high NCE and relatively high recovery of crop residues in animal rations.

## 5. Productivity frontier

The productivity frontier of agriculture represents maximum productivity and minimum N loss per agricultural hectare as a function of the N fixing ratio and N self-sufficiency. Figure S2 complements figure 2 of the main text by providing total N input per source and the share of vegetal and livestock in total N production and loss in the productivity frontier.

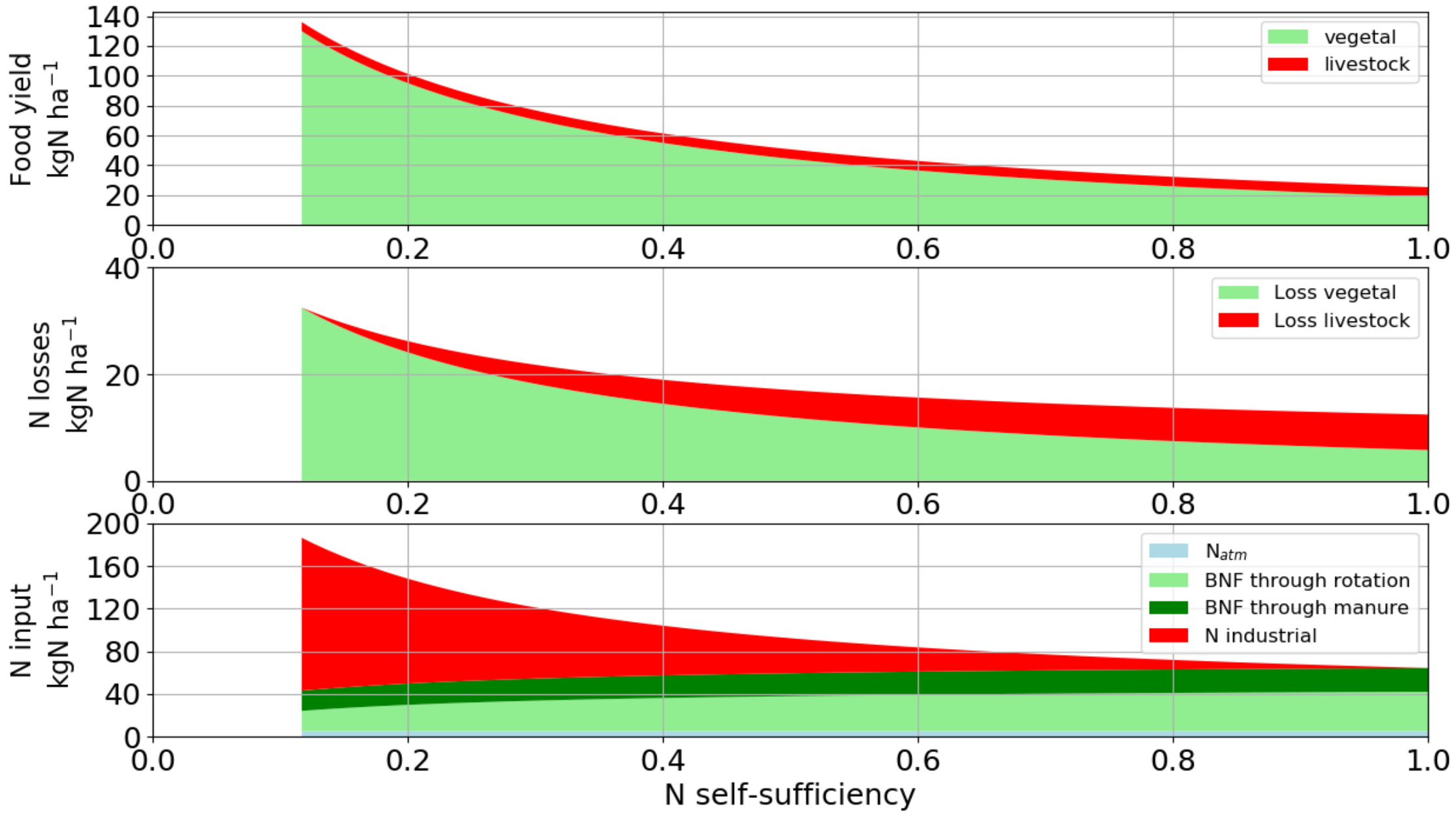

**Figure S2 | N food yield, N loss and total N input per source in function of the N fixing ratio and N self-sufficiency in the productivity frontier.** N self-sufficiency is calibrated for the case of manure N recycling of 0.7.

Table S2 summarizes all supporting input and output data allowing to position world regions in the productivity frontier. Feed-grain calculation is described in the Methods section *Feed calculation from FAOSTAT data* of the main text. The following section explains the calculation of N self-sufficiency and food yield gaps at the scale of world regions

**Table S2| Underlying data and model results for all world regions in 2013.** Sources: calculated [(1)] from FAOSTAT land and production data (see section 1) [(2)] from land data from FAOSTAT and N fixing rates from literature (Methods section *BNF calculation* of the main text) [(3)] based on FAOSTAT Food balance sheet [16] (Methods section *Feed calculation from FAOSTAT data* of the main text [(4)] based on *Mueller et al.* [17] (see section 6)

| | underlying data | | | | | | | | | results | | | | | |
|---|---|---|---|---|---|---|---|---|---|---|---|---|---|---|---|
| Regions | population | agricultural land (1) | N fixing ratio (1) | $r_{BNF}^{fixing}$ (2) | $r_{BNF}^{majorcrops}$ (2) | $N_{ind}$ (1) | $Yield_{majorcrops}$ (1) | feed-grain (3) | maximum achievable $Yield_{majorcrops}$ | apparent N food yield | apparent N self-sufficiency | N food yield corrected for feed trade | N self-sufficiency corrected for feed trade | region specific food yield gap | share of LV in food yield |
| | Mhab | Mha | % | kgN ha$^{-1}$ | kgN ha$^{-1}$ | kgN ha$^{-1}$ | kgN ha$^{-1}$ | % | kgN ha$^{-1}$ | kgN ha$^{-1}$ | % | kgN ha$^{-1}$ | % | kgN ha$^{-1}$ | % |
| Oceania | 38 | 390 | 93.2 | 5.9 | 4.1 | 59 | 40 | 65 | 77 | 2.6 | 73 | 2.1 | 73 | 6.7 | 29 |
| North America | 351 | 403 | 75.9 | 29.5 | 4.2 | 154 | 100 | 50 | 100 | 27.4 | 43 | 18.8 | 43 | 28.1 | 14 |
| Central America | 168 | 117 | 81.0 | 9.2 | 5.1 | 87 | 39 | 46 | 80 | 6.7 | 39 | 1.8 | 45 | 18.2 | 37 |
| South America | 408 | 621 | 87.4 | 28.5 | 7.1 | 87 | 50 | 52 | 63 | 16.8 | 74 | 7.4 | 74 | 9.9 | 12 |
| Caribbean | 43 | 10 | 55.9 | 18.6 | 9.7 | 29 | 18 | 41 | 28 | 8.6 | 46 | 0.1 | 60 | 13.4 | 40 |
| Western Europe | 191 | 46 | 45.1 | 21.3 | 2.1 | 174 | 118 | 70 | 129 | 48.6 | 12 | 26.4 | 14 | 76.9 | 40 |
| Eastern Europe | 294 | 234 | 51.9 | 11.0 | 1.4 | 47 | 52 | 58 | 85 | 18.4 | 33 | 15.9 | 33 | 44.4 | 21 |
| Northern Europe | 102 | 32 | 62.5 | 13.9 | 3.5 | 187 | 97 | 68 | 113 | 25.8 | 15 | 10.4 | 18 | 47.1 | 44 |
| Southern Europe | 153 | 54 | 48.4 | 13.2 | 1.8 | 86 | 54 | 69 | 72 | 17.5 | 17 | 1.8 | 21 | 39.4 | 44 |
| Eastern Africa | 378 | 345 | 80.6 | 10.0 | 6.1 | 10 | 23 | 17 | 66 | 4.6 | 88 | 4.6 | 88 | 15.9 | 11 |
| Middle Africa | 144 | 163 | 86.5 | 7.2 | 3.4 | 3 | 14 | 18 | 54 | 2.4 | 95 | 2.2 | 97 | 9.4 | 7 |
| Northern Africa | 217 | 167 | 78.3 | 8.4 | 2.6 | 50 | 30 | 41 | 42 | 5.6 | 46 | 2.0 | 53 | 10.1 | 27 |
| Southern Africa | 62 | 158 | 96.4 | 6.5 | 2.0 | 78 | 50 | 44 | 89 | 1.7 | 78 | 1.3 | 80 | 4.2 | 41 |
| Western Africa | 334 | 300 | 65.1 | 9.3 | 5.8 | 7 | 15 | 27 | 36 | 4.8 | 83 | 4.6 | 84 | 14.5 | 9 |
| Central Asia | 66 | 282 | 90.9 | 5.7 | 0.8 | 31 | 35 | 56 | 70 | 2.1 | 78 | 2.0 | 78 | 8.2 | 29 |
| Eastern Asia | 1,620 | 698 | 75.3 | 9.7 | 8.9 | 183 | 70 | 49 | 86 | 14.6 | 22 | 7.2 | 24 | 24.3 | 27 |
| Southern Asia | 1,777 | 364 | 34.1 | 41.4 | 12.5 | 95 | 38 | 37 | 51 | 20.7 | 30 | 20.1 | 30 | 35.7 | 20 |
| South-Eastern Asia | 620 | 143 | 16.5 | 35.5 | 15.9 | 63 | 32 | 43 | 35 | 22.6 | 29 | 14.6 | 32 | 31.0 | 15 |
| Western Asia | 248 | 260 | 88.2 | 6.0 | 2.8 | 76 | 47 | 51 | 78 | 4.4 | 47.5 | 1.6 | 54 | 11.5 | 29 |
| World | 7200 | 4787 | 74.3 | 15.0 | 7.5 | 88 | 47 | 46.0 | 70 | 9.4 | 43.9 | ND | ND | 20.4 | 26 |

## 6. Regional N self-sufficiency and food yield gaps in the productivity frontier

We estimate food yield and N self-sufficiency per world region in 2013 based on FAOSTAT data [2]. At regional scales, net feed trade ($N_Feed_{net-imp}$) i.e. the difference between imports and exports, is a significant component of N budgets but trade data do not report the share of feed in total trade [18]. We calculate $N_Feed_{net-imp}$ per world region with equation 10 by assigning to each main crop category *i* the world average share of feed in total crop production of that category (Table S3). Oilcrops processed into oil and cake are accounted for by allocating all N supply to cake and all cake to feed [18].

$$N_Feed_{net-imp} = Max(0; N_Feed_{imp} - N_Feed_{exp}) \text{ with } N_Feed_{imp} - N_{Feed}{}_{exp} = \sum_i \gamma_i \times \left(N_{imp}^i - N_{exp}^i\right) \qquad \textit{(eq. 10)}$$

where $\gamma_i$ is the global average share of crop *i* used as feed in production and $N_{exp}^i$ , $N_{imp}^i$ are respectively the N exports and imports for crop *i* (ref. [9,16]).

**Table S3| World average share of feed per main crop category in 2013,** calculated from FAOSTAT [2].

| Product | Pulses | Oilcrops seed and cake | Maize and products | Sorghum and products | Wheat and products | Rice and products | Millet and products | Cereals, Other | Barley and products | Oats |
|---|---|---|---|---|---|---|---|---|---|---|
| Share of feed $\gamma_i$ | 18% | 86% | 61% | 48% | 39% | 34% | 27% | 66% | 69% | 76% |

Each region is positioned twice. Once, in terms of apparent food yield and N self-sufficiency and once after correction for net feed trade. The apparent food yield ($Food\ yield_{\text{apparent}}$) is the biomass output of agriculture regardless of the potential dependence of production on net feed imports and of the nature (food or feed) of the biomass exports. The correction for feed trade is done by subtracting net feed imports from domestic production and by accounting for net feed exports in terms of food equivalent. The food equivalent of feed is calculated by multiplying net feed exports by the world average N conversion efficiency of livestock (NCE). The corrected food yield ($Food\ yield_{\mathrm{corrected}}$) is a measure of the true contribution of a region in world food availability. The apparent and corrected Food yield and N self-sufficiency are calculated with equations 11 to 14. They are provided per world region in Table S2 (columns 10 to 15).

$$Food\ yield_{\mathrm{apparent}} = N_Prod_{VL} + N_Prod_{LV} - N_Feed_{total} + N_Feed_{net-imp} \quad \text{(eq. 11)}$$

$$N_{self-sufficiency\ apparent} = (BNF + N_{atm})/(BNF + N_{atm} + N_{ind} + N_{Feed_{net-imp}}) \quad \text{(eq. 12)}$$

$$Food\ yield_{\mathrm{corrected}} = N_Prod_{VL} + N_Prod_{LV} - N_Feed_{total} - (1 - NCE) \times Max(0; N_Feed_{exp} - N_Feed_{imp}) \quad \text{(eq.13)}$$

$$N_{self-sufficiency\ corrected} = (BNF + N_{atm})/(BNF + N_{atm} + N_{ind}) \quad \text{(eq.14)}$$

where $N_Prod_{VL}$ is the regional crops production calculated from FAOSTAT Production Crops module using N content factors [16,19], $N_Prod_{LV}$ is the indigenous livestock primary production excluding fish production calculated from FAOSTAT Livestock Primary indigenous Production module (defined as indigenous animals slaughtered, plus the exported live animals of indigenous origin, see FAO definition [20], $N_{ind}$ is industrial N fertilizer use (FAOSTAT Fertilizer module), $N_Feed_{total}$ is the N feed calculated for all crops provided in FAO Food Balance Sheet (FBS) [18] as described in the Methods section *Feed calculation from FAOSTAT data* of the main text, $N_Feed_{net-imp}$ is net N feed imports calculated above and BNF per world region calculated in section 2 (Table S2).

For each region, current major crops yield is calculated as the weighted average yield of all crops in major crops land based on FAOSTAT crops production data in 2013 (Table S2 in column 8). It is compared to maximum major crops yield by assuming a full closure of yield gaps in each region [17]. The calculation is based on the yield gaps calculated for thirteen main crops in *Mueller et al.* [17] considering the crop mix in 2013 in each region. For crops not reported in the yield gap analysis in *Mueller et al.* [17], we applied the yield gap ratio for reported crops on the yields in 2013 of non-reported crops (Table S2, column 10, the country list per region is in Table S4).

Regional food yield gaps are quantified by taking into account region-specific crop yield gaps. Region specific maximum food yields are highlighted on the line connecting each region to the productivity frontier according to the regional N fixing ratio. The analysis highlights that the food yield is sensibly lower than the productivity frontier in most world regions.

**Table S4| List of countries by region based on FAO categorization**

| Oceania | North America | Central America | South America | Caribbean | Western Europe | Eastern Europe | Southern Europe | Northern Europe | Eastern Africa | Middle Africa | Northern Africa | Southern Africa | Western Africa | Western Asia | Central Asia | Eastern Asia | Southern Asia | South-Eastern Asia |
|---|---|---|---|---|---|---|---|---|---|---|---|---|---|---|---|---|---|---|
| Australia | Canada | Belize | Argentina | Antigua and Barbuda | Austria | Belarus | Albania | Denmark | Burundi | Angola | Algeria | Botswana | Benin | Armenia | Kazakhstan | China, Hong Kong SAR | Afghanistan | Brunei Darussalam |
| Fiji | United States of America | Costa Rica | Bolivia | Bahamas | Belgium | Bulgaria | Bosnia and Herzegovina | Estonia | Comoros | Cameroon | Egypt | Lesotho | Burkina Faso | Azerbaijan | Kyrgyzstan | China, mainland | Bangladesh | Cambodia |
| Guam | | El Salvador | Brazil | Barbados | France | Czechia | Croatia | Finland | Djibouti | Central African Republic | Libya | Namibia | Cabo Verde | Cyprus | Tajikistan | China, Taiwan Province of | Bhutan | Indonesia |
| Micronesia (Federated States of) | | Guatemala | Chile | Cuba | Germany | Hungary | Greece | Ireland | Eritrea | Chad | Morocco | South Africa | Côte d'Ivoire | Georgia | Turkmenistan | Democratic People's Republic of Korea | India | Lao People's Democratic Republic |
| New Caledonia | | Honduras | Colombia | Dominica | Luxembourg | Poland | Italy | Latvia | Ethiopia | Congo | Sudan | Swaziland | Gambia | Iraq | Uzbekistan | Japan | Iran (Islamic Republic of) | Malaysia |
| New Zealand | | Mexico | Ecuador | Dominican Republic | Netherlands | Republic of Moldova | Malta | Lithuania | Kenya | Dem. Republic of the Congo | Tunisia | | Ghana | Israel | | Mongolia | Maldives | Myanmar |
| Papua New Guinea | | Nicaragua | French Guiana | Grenada | Switzerland | Romania | Montenegro | Norway | Madagascar | Gabon | Western Sahara | | Guinea | Jordan | | Republic of Korea | Nepal | Philippines |
| Solomon Islands | | Panama | Guyana | Haiti | | Russian Federation | Portugal | Sweden | Malawi | Sao Tome & Principe | | | Guinea-Bissau | Kuwait | | | Pakistan | Thailand |
| Vanuatu | | | Paraguay | Jamaica | | Slovakia | Serbia | United Kingdom | Mauritius | | | | Liberia | Lebanon | | | Sri Lanka | Timor-Leste |
| | | | Peru | Montserrat | | Ukraine | Slovenia | | Mozambique | | | | Mali | Occupied Palestinian Territory | | | | Viet Nam |
| | | | Suriname | Puerto Rico | | | Spain | | Réunion | | | | Mauritania | Oman | | | | |
| | | | Uruguay | Trinidad and Tobago | | | The for. Yugoslav Republic of Macedonia | | Rwanda | | | | Niger | Qatar | | | | |
| | | | Venezuela | | | | | | Somalia | | | | Nigeria | Saudi Arabia | | | | |
| | | | | | | | | | South Sudan | | | | Senegal | Syrian Arab Republic | | | | |
| | | | | | | | | | Uganda | | | | Sierra Leone | Turkey | | | | |
| | | | | | | | | | United Republic of Tanzania | | | | Togo | United Arab Emirates | | | | |
| | | | | | | | | | Zambia | | | | | Yemen | | | | |
| | | | | | | | | | Zimbabwe | | | | | | | | | |

## 7. Agricultural land in the boundaries

All food production boundaries consider constant total agricultural land as in 2013 (Table S1). This assumption is based on planetary system sustainability analysis considering interactions among various planetary boundaries defined for different resources including land use and associated biomes [21]. The moderate expansion in agricultural land over the last decades shown in Figure S1 has mainly occurred at the expenses of natural ecosystems, in particular forestland in the tropics [22], with high ecological value and subsequent environmental impacts in terms of greenhouse gas emissions, biodiversity loss, soil degradation and water quality [23,24]. Recent studies on prospective agricultural scenarios [11,25,26] have often considered a net expansion in total agricultural land, before concluding that such an expansion would be incompatible with the respect of planetary boundaries and environmental sustainability goals. If forestland were to be preserved for environmental sustainability, the assumption of constant world agricultural land as in 2013 is most probably optimistic. This is due to ongoing soil degradation relating to unsustainable intensification of production practices in many world regions. For example, some 1.5 million ha of intensively irrigated arable land (0.1 % of total cropland) are estimated to be heavily salinized each year and risk to be abandoned worldwide [22].

## 8. Current situation and historical trends of $NUE_{crop}$ and yields

Figure S3 shows $NUE_{crop}$ and crop yields for all countries in 2011 as calculated in *Lassaletta et al*. [9]. Countries will $NUE_{crop}$ above 1 operate soil N mining [9]. The five countries framed in red in figure S3 (Austria, France, Germany, Denmark, USA) display major crops yield and $NUE_{crop}$ close to the values considered in the world food production boundaries. For these five countries, as well as for China (the largest crop producer) and the world, we show in figure S4 the historical trajectories of major crops yield and $NUE_{crop}$ since 1961 based on *Lassaletta et al.* [9] and *Zhang et al.* [27]. The five best performing countries have managed to increase simultaneously the major crops yield and the $NUE_{crop}$ at least over a sub-period since 1961. For China, the crop yield continues to increase at the expense of the $NUE_{crop}$. For the world, the crop yield continues to increase while the $NUE_{crop}$ tends to stabilize. In each panel, the calculated annual growth rates and the related time period are used in Figure 4 of the main text.

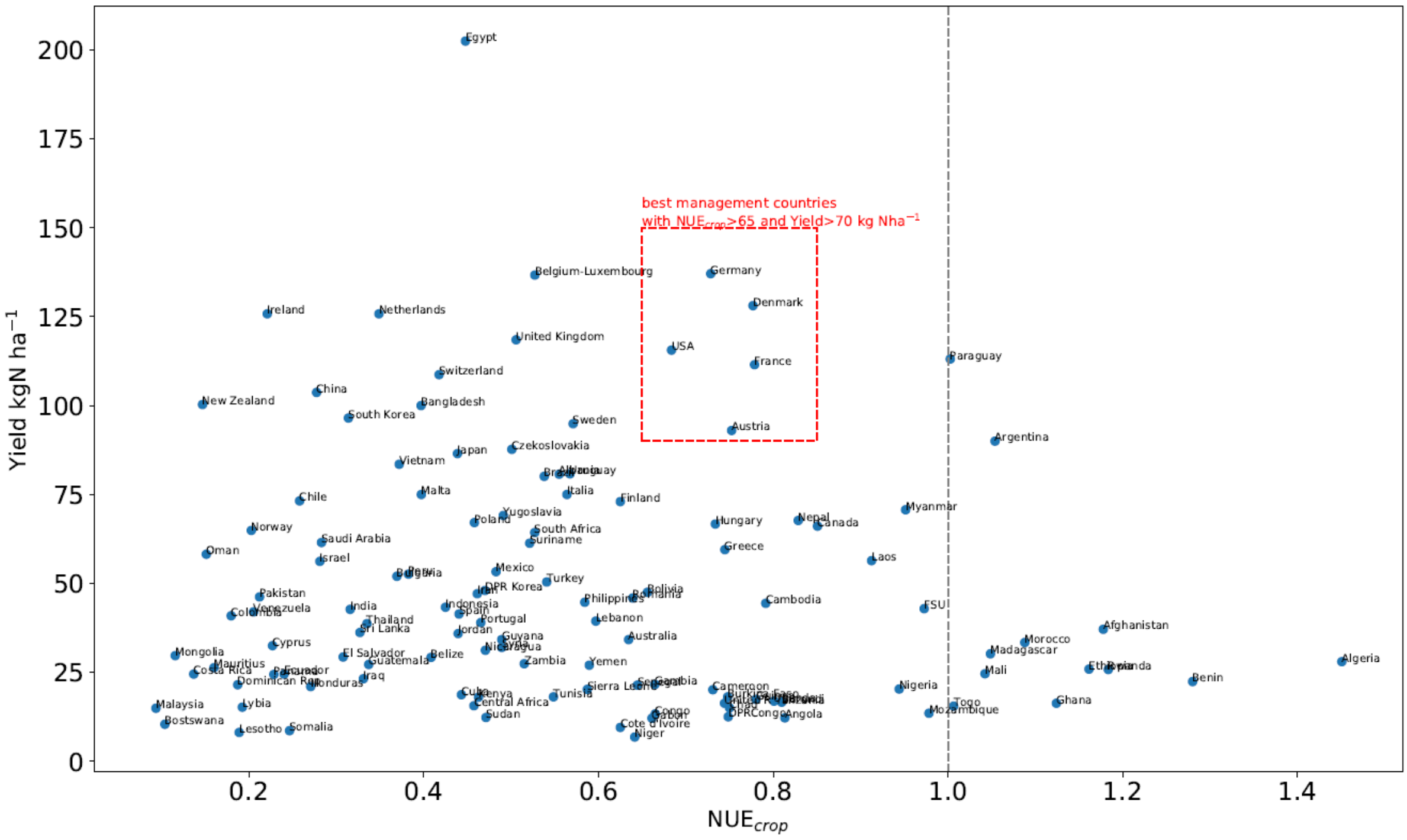

**Figure S3| Major crops yield and $NUE_{crop}$ per country in 2011 based on *Lassaletta et al*.** [9]. We framed the five countries (Austria, France, Denmark, Germany and United States of America) that combine high $NUE_{crop}$ and yield (respectively above 65% and 70 kgN ha$^{-1}$) close to the values considered in world food production boundaries. Countries with $NUE_{crop}$ above 1 operate soil N mining.

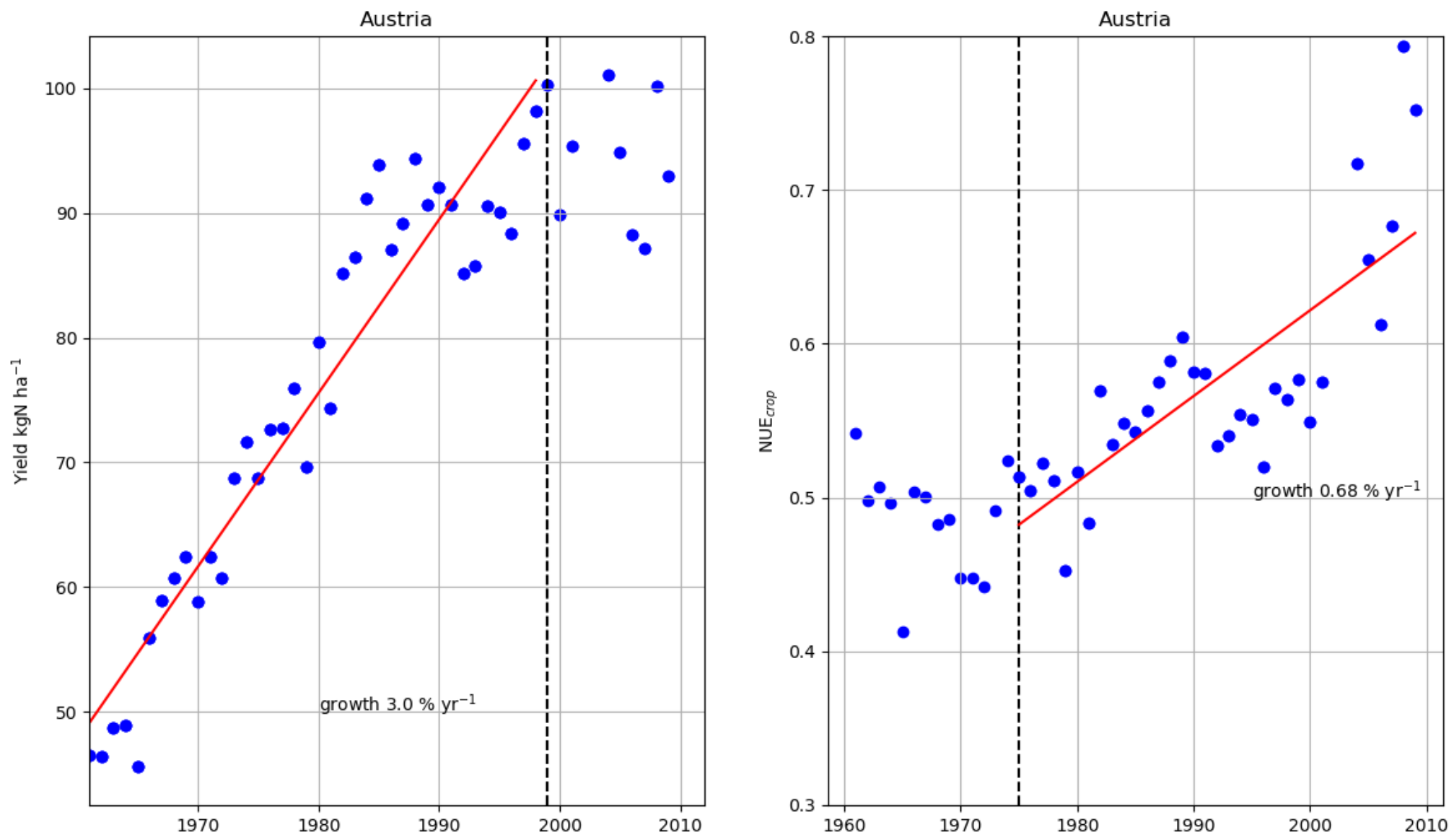

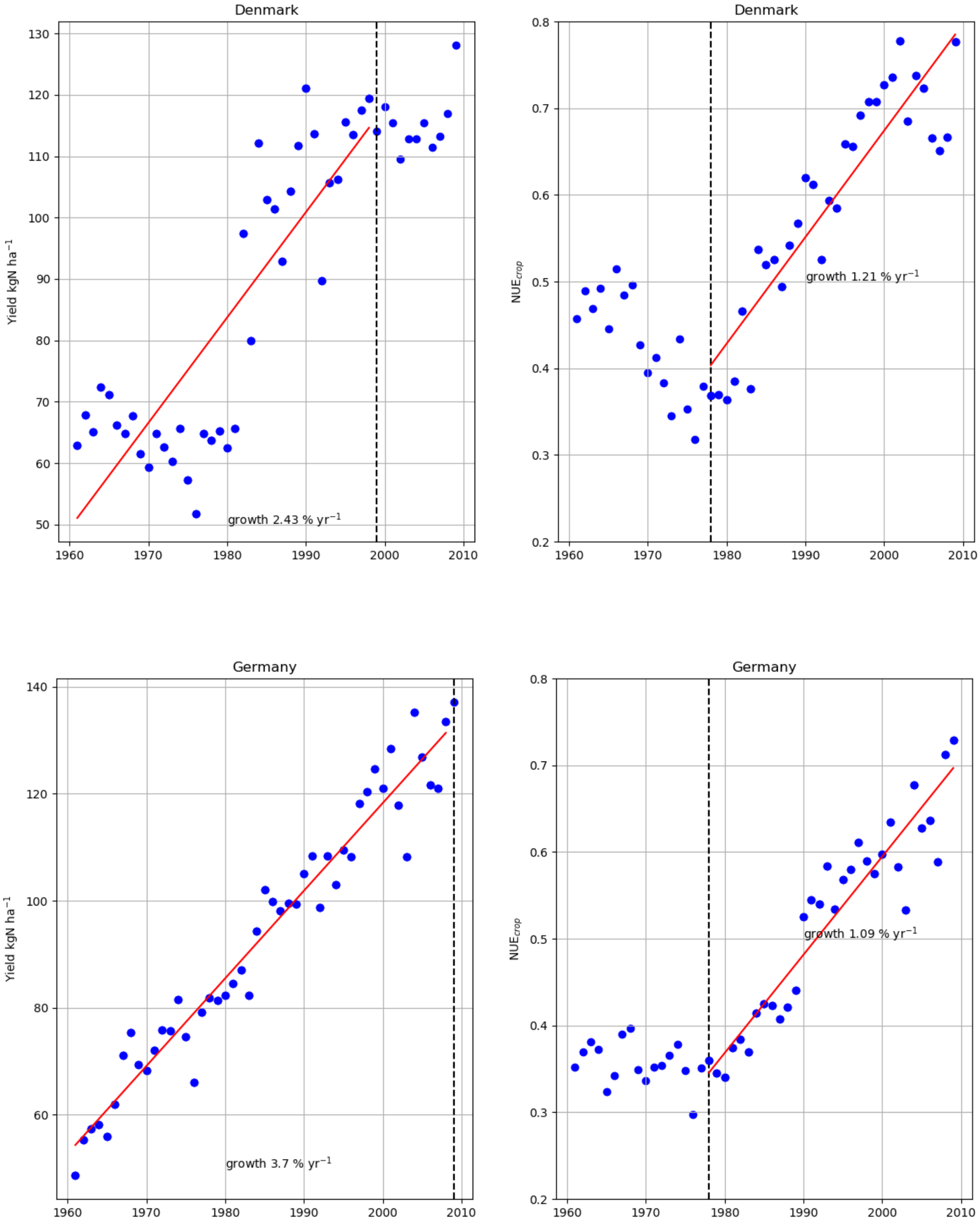
Denmark
Yield kgN ha$^{-1}$
growth 2.43 % yr$^{-1}$
Denmark
NUE$_{crop}$
growth 1.21 % yr$^{-1}$
Germany
Yield kgN ha$^{-1}$
growth 3.7 % yr$^{-1}$
Germany
NUE$_{crop}$
growth 1.09 % yr$^{-1}$

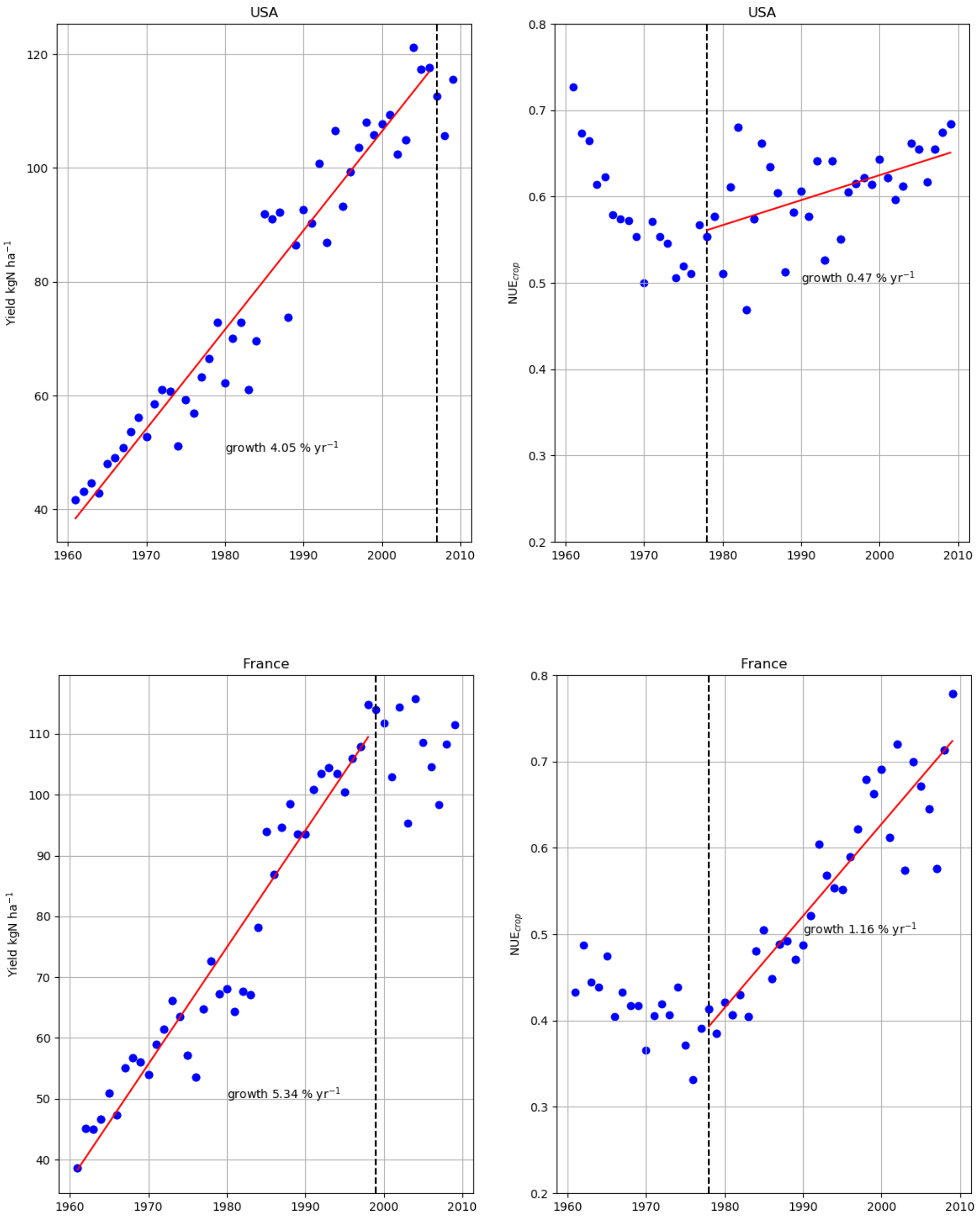
USA
Yield kgN ha$^{-1}$
growth 4.05 % yr$^{-1}$
USA
$NUE_{crop}$
growth 0.47 % yr$^{-1}$
France
Yield kgN ha$^{-1}$
growth 5.34 % yr$^{-1}$
France
$NUE_{crop}$
growth 1.16 % yr$^{-1}$

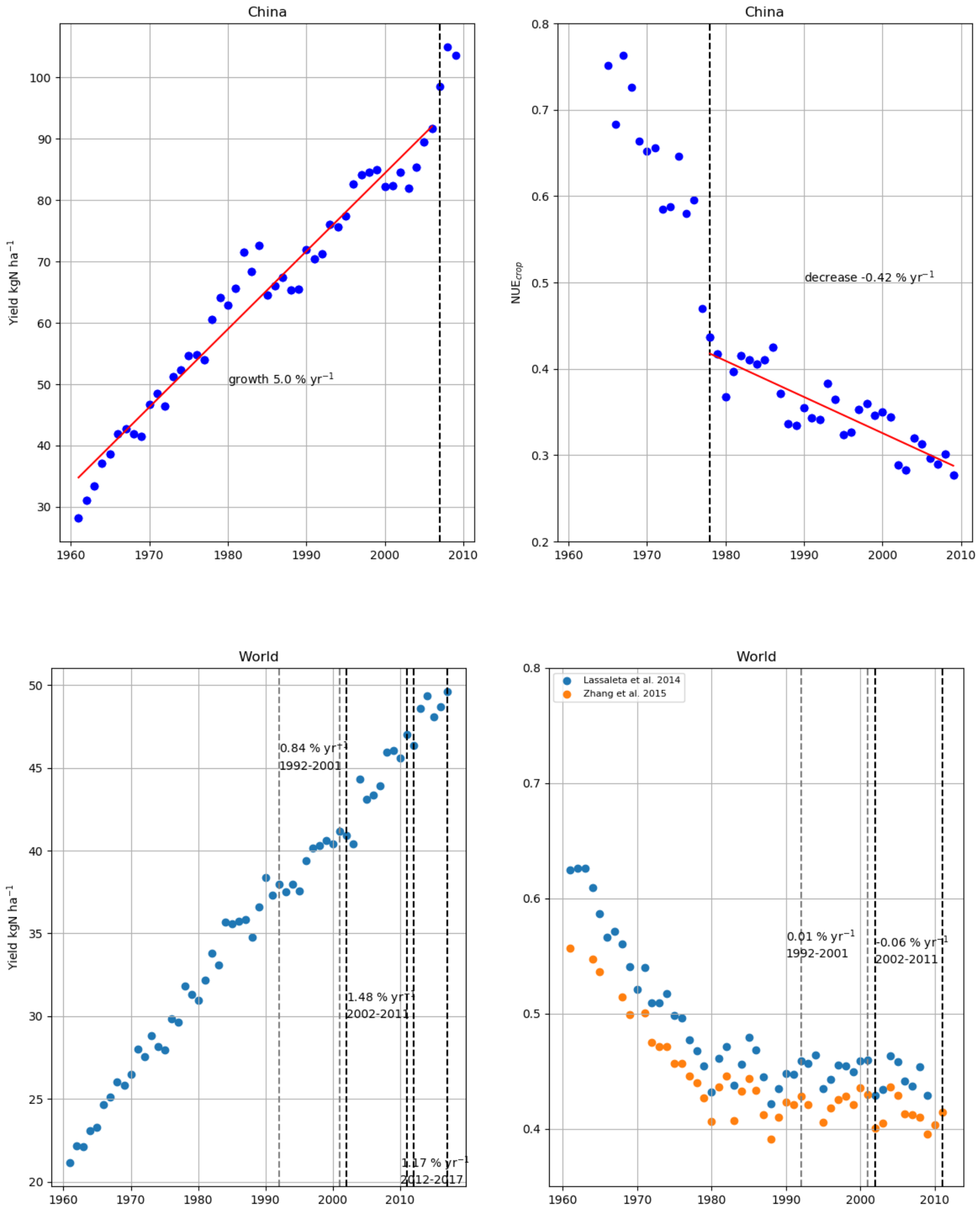

**Figure S4| Historical crop yields and $NUE_{crop}$ for the five best performing countries (Austria, Germany, France, Denmark and USA), China as the largest crop producer and the world.** Country level crop yields and $NUE_{crop}$ are based on *Lassaletta et al*. [9]. World major crop yields are calculated from FAOSTAT (2018) and the associated $NUE_{crop}$ are derived based on *Zhang et al.* [27] and *Lassaletta et al*. [9]. The vertical dotted lines define the time period used in the calculation of the growth rates in major crops yield and $NUE_{crop}$ in figure 4 of the main text.

## 9. Yield and N loss calculation in organic boundaries

Figure S5 shows the variation in major crops yield, N loss and N food yield in B4a (fig S5a) and B4b (fig S5b) as a function of the variation in N fixing land ($\Delta S_{fixing}$) with current feed-grain. The maximum food yield is found for $\Delta S_{fixing}$ of 470 Mha in B4a and of about 200 Mha in B4b but the increase is quite small in both cases. $\Delta S_{fixing}$ is limited to 100 Mha in both B4a and B4b.

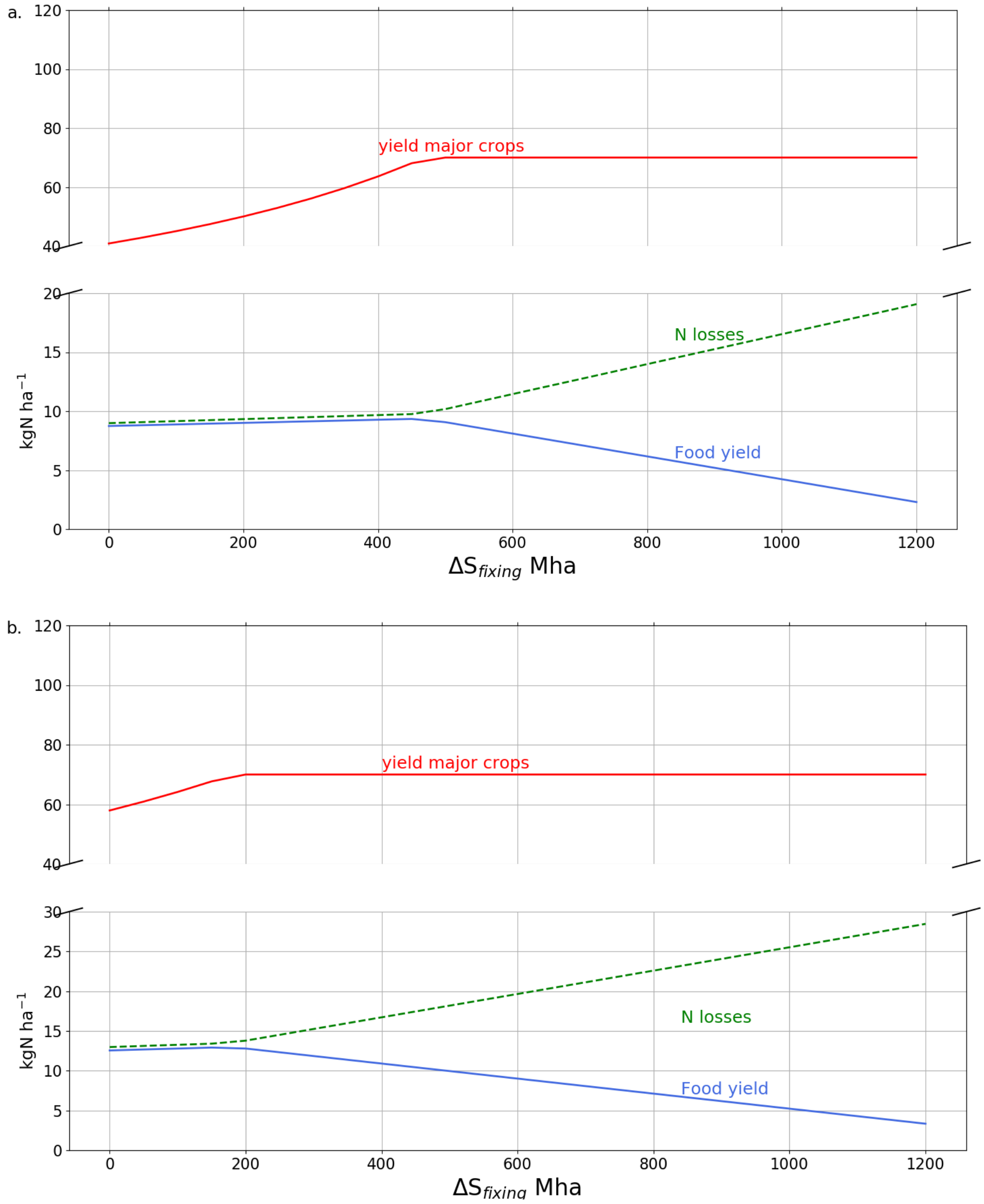

**Figure S5| Major crops yield, N loss and N food yield in B4a (a) and B4b (b) as a function of the variation in N fixing land ($\Delta S_{fixing}$>0) with current feed-grain.** N food yield maximizes for $\Delta S_{fixing}$ of 470Mha in B4a and of about 200 Mha in B4b. Note that major crops yield and the N food yield maximize for slightly different $\Delta S_{fixing}$.

## 10. N planetary boundary thresholds

The N planetary boundary indicates the range of allowable N loss from agriculture to remain within the safe N operating space. It is calculated based on the range of 50 to 100 TgN provided in *Bodirsky et al.* [12] considering thresholds analysis of acceptable air, water and atmospheric pollution [28]. When expressed per unit of total agricultural land, the allowable N loss from agriculture ranges from 10.5 to 21 kgN $ha^{-1}$ $yr^{-1}$. These values define the lower and upper limit of the N planetary boundary in figure 4 of the main text.

## 11. Difference between N food yield, N household food supply and the recommended N intake

Based on FAO FBS [18] and the N balance calculation (described in the Methods section *Feed calculation from FAOSTAT data* of the main text), world average food yield per capita in 2013 is estimated at 6.10 kgN $cap^{-1}$ $yr^{-1}$, out of which 1.50 kgN $cap^{-1}$ $yr^{-1}$ is livestock production. Average household food supply from terrestrial ecosystems (excluding 0.3 kgN $cap^{-1}$ $yr^{-1}$ of fish and seafood) is 4.45 kgN $cap^{-1}$ $yr^{-1}$ (ref. [18]). The recommended N intake ranges from 3 to 4 kgN $cap^{-1}$ $yr^{-1}$ (ref [29]). Accordingly, world average food yield per capita is almost twice the recommended N intake. The difference of about 0.45-1.35 kgN $cap^{-1}$ $yr^{-1}$ between household food supply and the recommended N intake is overconsumption for a significant part of the world population and household waste.

The difference between the food yield and food supply is almost 1.65 kgN $cap^{-1}$ $yr^{-1}$ and corresponds to non-food uses (35%), seed (15%) and loss (50%). This difference amounts to 35% of the share of vegetal production in food yield and is kept constant in the boundaries.

Figure S6a shows the cumulative distribution of household food supply over the world population in 1961 and 2013, highlighting that the share of global population receiving less dietary N than recommended has drastically decreased. Figure S6b shows the distribution of N food supply per capita and country according to the World Bank classification of countries based on income. Most of the high-income countries are in the upper part of the N supply distribution and most of the low-income countries are in the lower part. This distribution supports the view that N loss increases with income and that the economic transition in currently low-income countries might exacerbate global N loss.

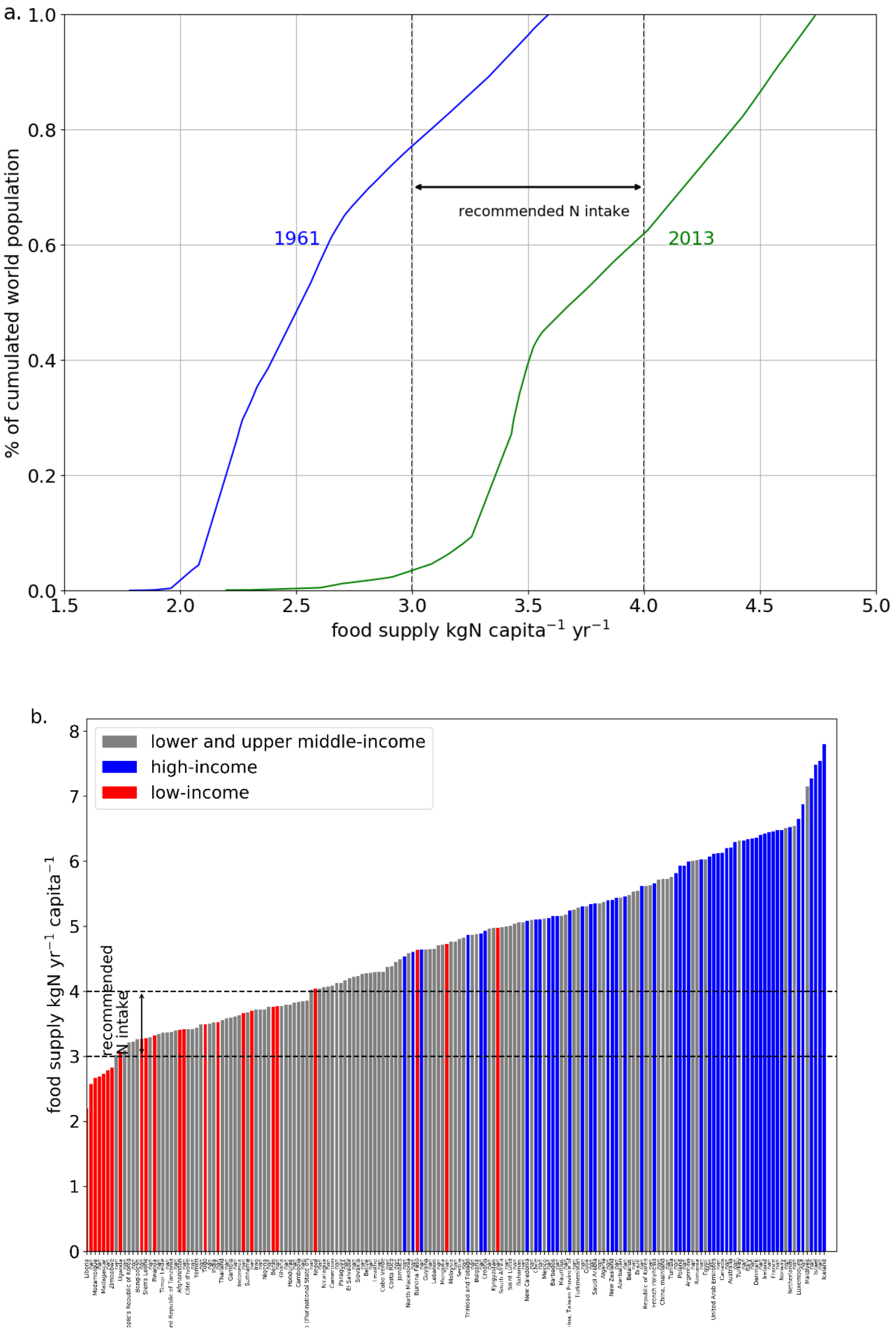

**Figure S6| a. N food supply distribution over the world population in 1961 and 2013 b. N food supply in 2013 per country**. Countries are grouped in three categories according to the World Bank classification by income[30]. The range of recommended N intake (horizontal dotted lines) is calculated assuming a protein intake of 1g kg$^{-1}$ day$^{-1}$ (0.83g being the safe level for adult men and women [29] and a global average human body weight ranging between 50 and 70 kg).

## 12. The world trajectory of feed-grain

Based on the N balances constructed (Methods section *Feed calculation from FAOSTAT data* of the main text), we calculate the share and total mass of grain allocated to feed (feed-grain) at the global scale since 1961 (figure S7). Feed-grain has varied between 45 and 52 % since 1961 and is at 46 % today. However, besides the reduction in the share of feed-grain, the mass of feed-grain has tripled since 1961 following the increase by 2.3 in major crops yield and the increase by 30% in major crops land.

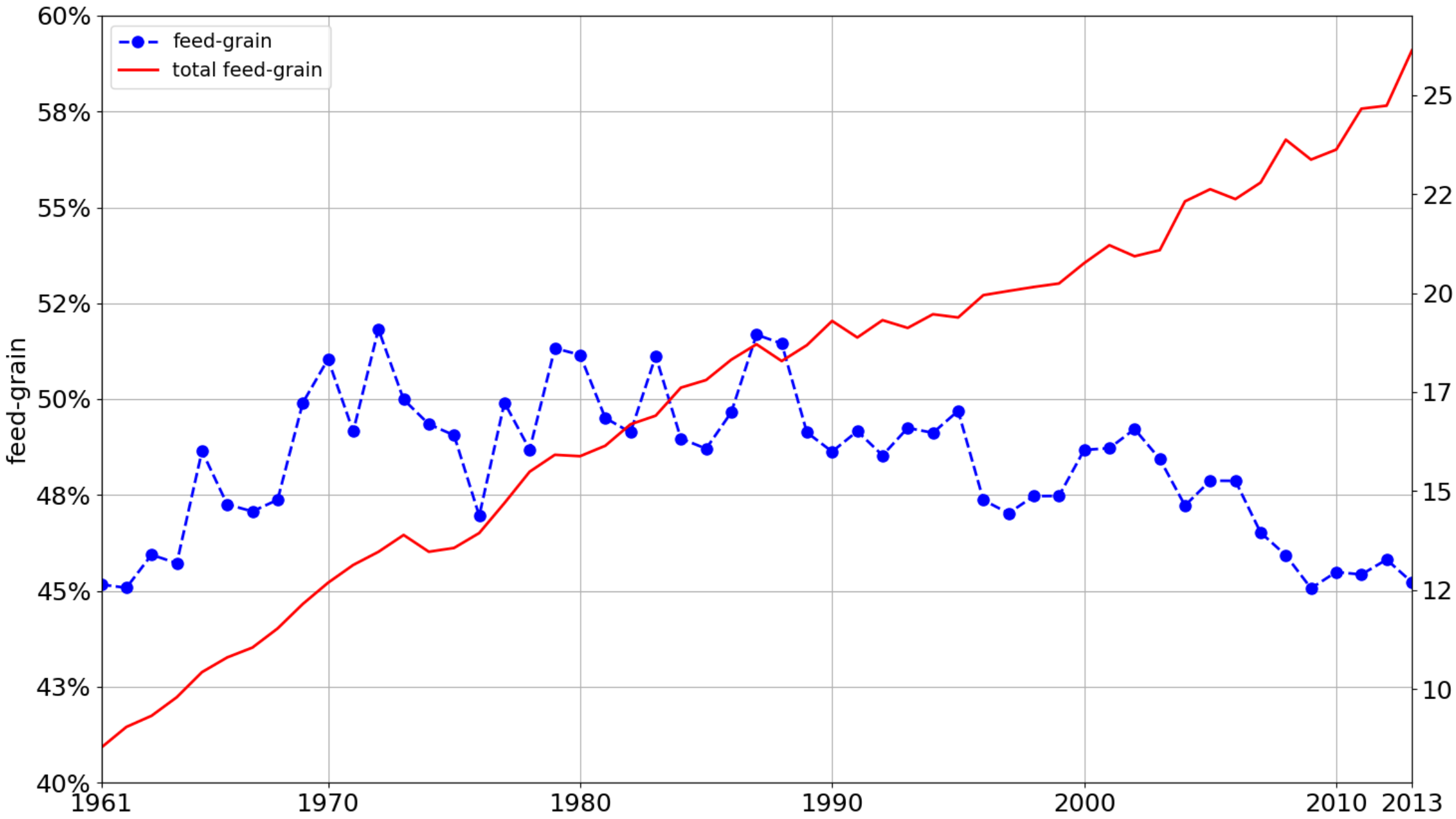

**Figure S7| Historical world average feed-grain in % and Tg N between 1961 and 2013.**

## 13. Livestock production in food yield, $NUE_{tot}$ and N input per source

Figure S8 shows the share of livestock production in food yield and food supply in the different boundaries for three values of feed-grain (0, 0.46 and 0.7). 70 % is the share currently at stake in Western Europe, 46 % is the world average and 0 is a theoretical value maximizing agricultural productivity. The share of livestock production is higher in the food supply than in the food yield because of the share of vegetal production going to non-food uses, seed and loss ([2] and section 11).

$NUE_{tot}$ maximizes for maximum manure N recycling rate and $NUE_{crop}$ and minimum feed-grain (zero). Figure S9 shows in detail the sensitivity of $NUE_{tot}$ to these three variables in the case of $N_{ind}$-based boundaries (B1, B2, B3). For a given level of manure N recycling rate, the higher the feed-grain, the lower the sensibility of $NUE_{tot}$ to $NUE_{crop}$ (Figure S9a). For a given level of $NUE_{crop}$, the higher the manure N recycling rate, the lower the impact of feed-grain on $NUE_{tot}$ (Figure S9b).

Figure S10 shows $N_{tot}$ per source in time and in the five food production boundaries. In $N_{ind}$-based boundaries (B1, B2, B3), we made the simplifying assumption that all $N_{ind}$ is applied to major crops land. According to the International Fertilizer Association[31]**,** 4.3% of the world total $N_{ind}$ is applied to grasslands and 5.4% to soybean in 2014.

With current feed-grain, $N_{ind}$ in B1 is 160 TgN which is about 60% higher than today (105 TgN in 2013) and in agreement with *Conijn et al.* [11] who estimate $N_{ind}$ to feed 9.7 billion people with current diets and unimproved $NUE_{crop}$. $N_{ind}$ is the lowest in B3 amounting to 38 TgN. This figure is higher than the estimate of 27 TgN in *Bodirsky et al.* [12] regarding a scenario with 9.1 billion people in 2050, improved

N efficiency in agriculture, N recovery from waste and about 15% less livestock-based calories in human diets.

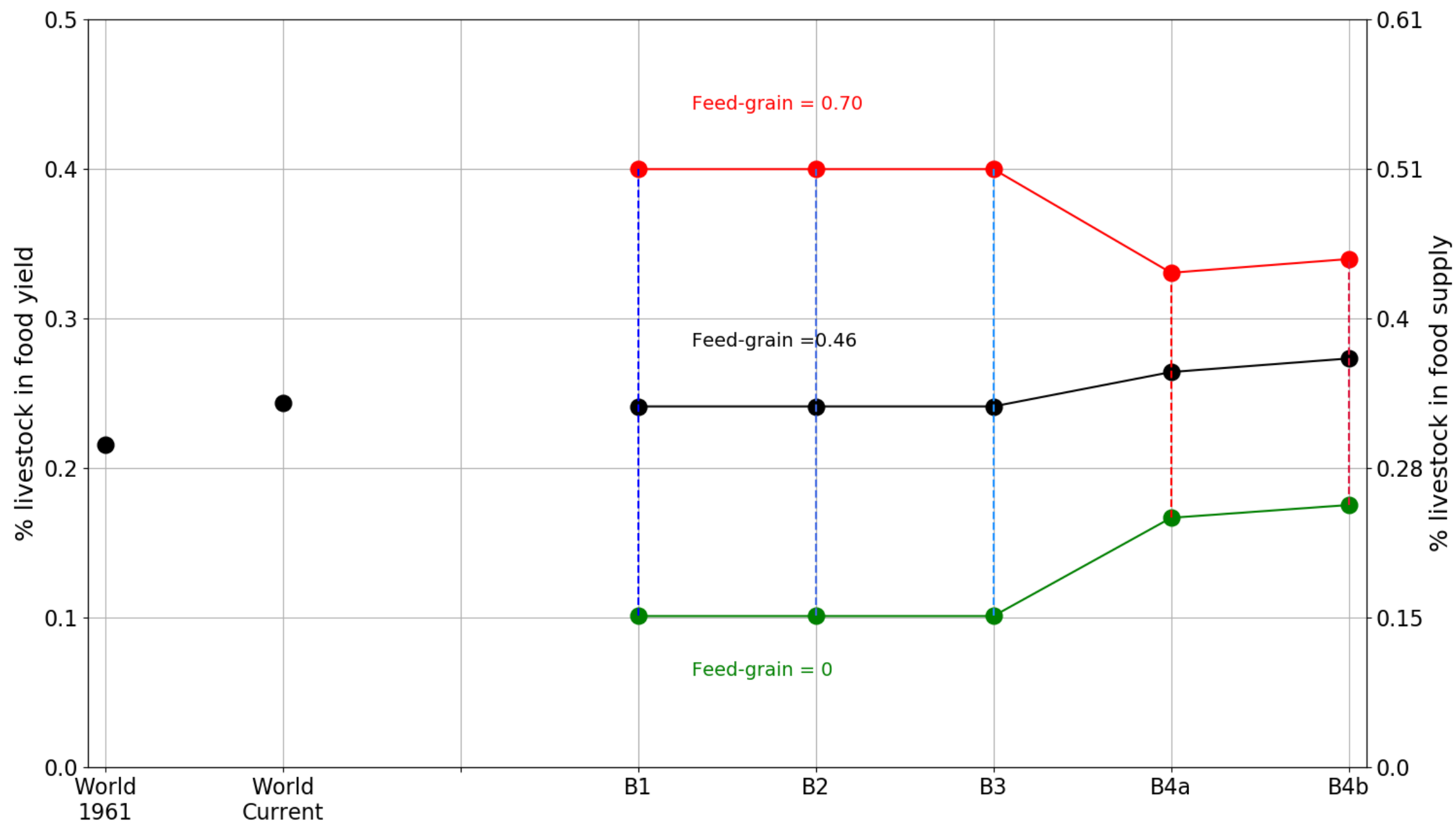

**Figure S8| Share of livestock in food yield (left axis) and food supply (right axis) in 1961 and 2013 and in the five food production boundaries (B1, B2, B3, B4a, B4b) with feed-grain equal to 0.46 (current value), 0 and 0.7.**

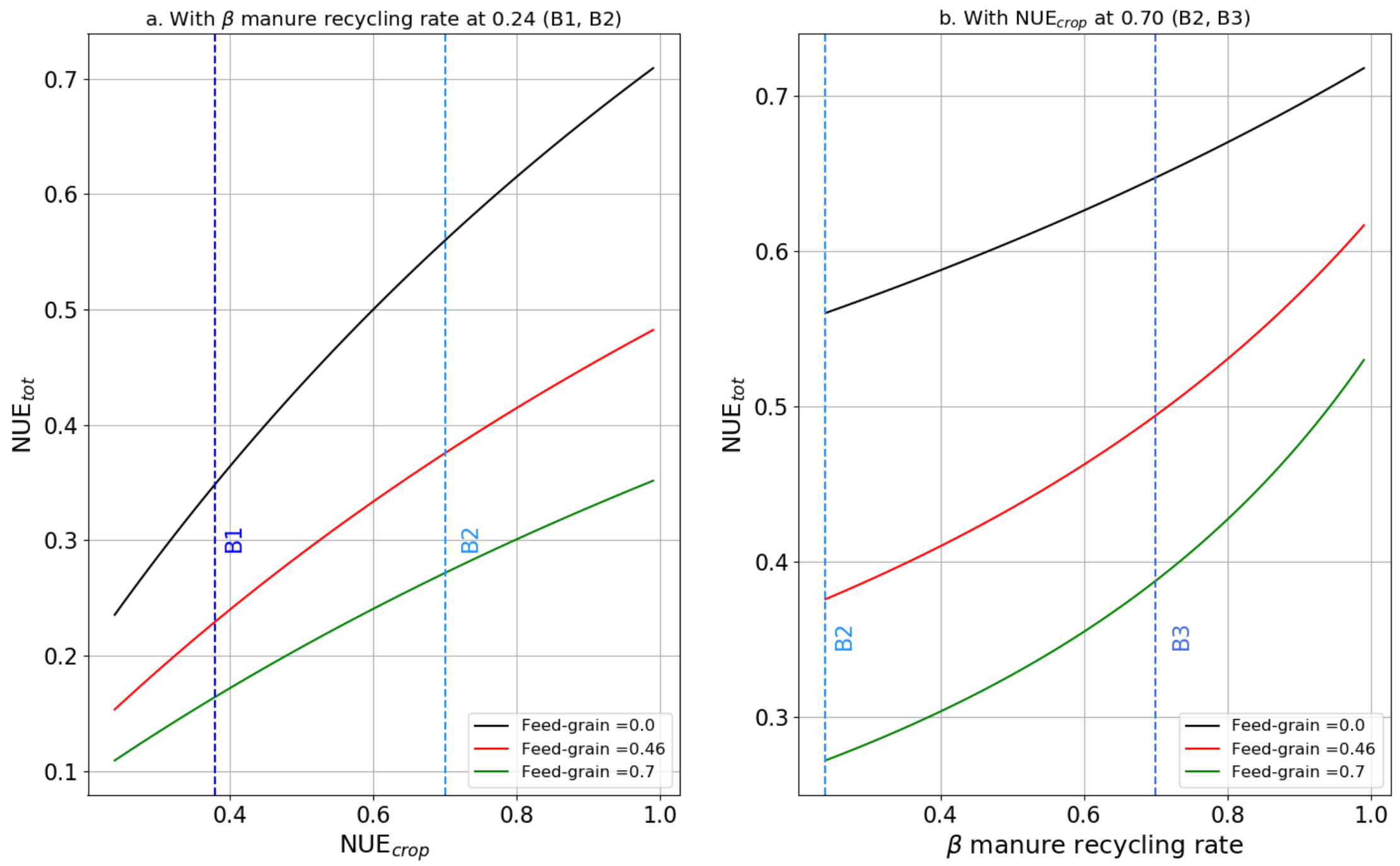

**Figure S9| $NUE_{tot}$ sensibility to $NUE_{crop}$, manure N recycling rate (β) and feed-grain. a. $NUE_{tot}$ in function of $NUE_{crop}$ with β at 0.24 as today (B1, B2) b. $NUE_{tot}$ in function of β and $NUE_{crop}$ at 70% (B2, B3).** In both figures, $NUE_{tot}$ is calculated with feed-grain equal to 0.46 (current value), 0 and 0.7.

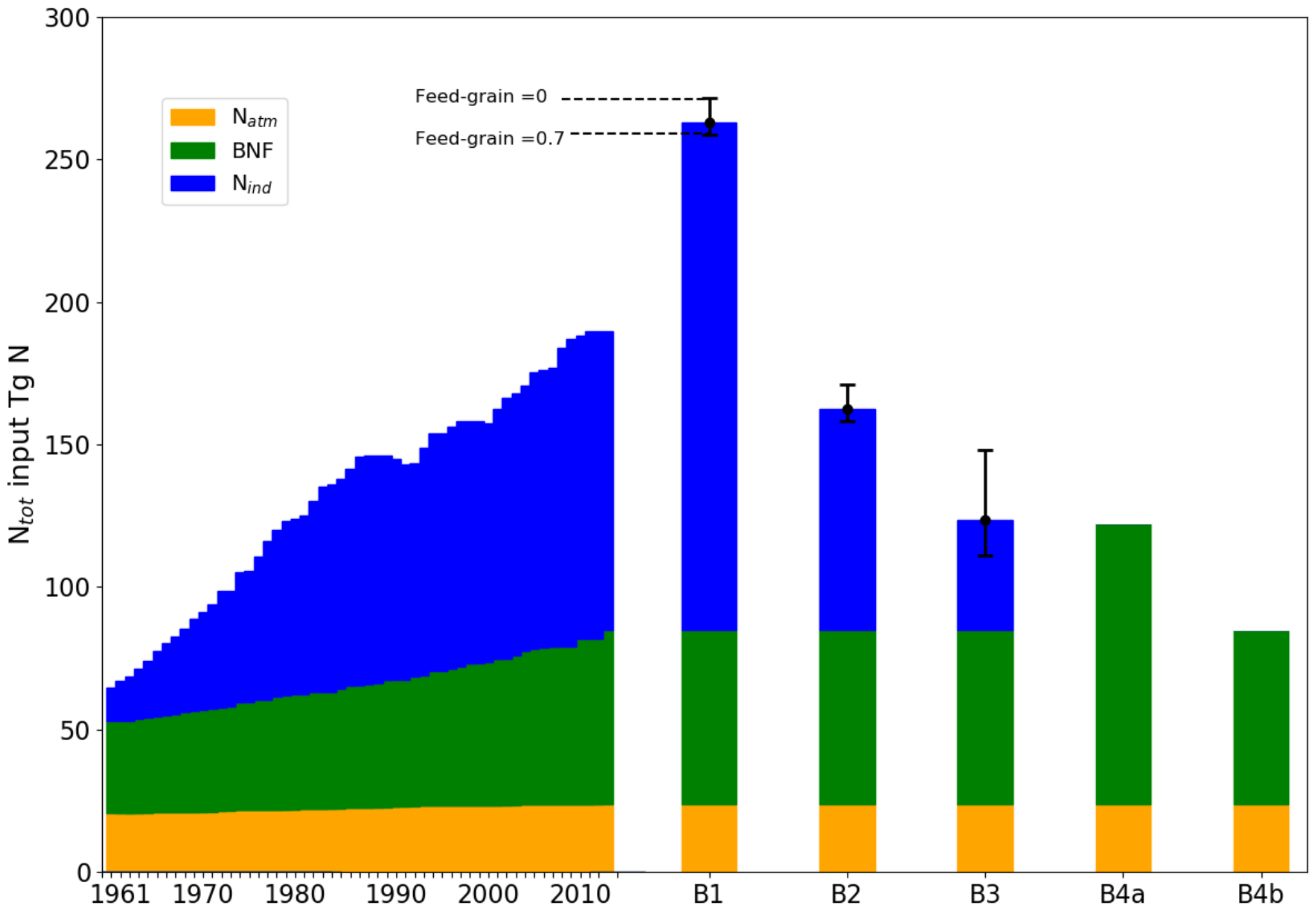

**Figure S10| $N_{tot}$ per N source from 1961 to 2013 and in the five food production boundaries with feed-grain equal to 0.46 (current value).** The effect of feed-grain on $N_{ind}$ is indicated by the vertical bars with feed-grain equal to 0 and 0.7.

**Table S5| Calculation of the average annual growth rates (% $yr^{-1}$) required in the key agricultural variables governing production and efficiency to achieve both the food production boundaries and the N planetary boundary (B3, B4b) by 2050 with the current share of feed-grain.** All rates are non-compounding. For NCE, manure N recycling and $NUE_{crop}$ which are percentages, the growth rates are calculated as the simple ratio of difference between current and target values over the number of years. For organic agriculture, the growth rate is calculated by considering a reference yield of 38, which is 20% below the global average[32].

| | Variables | Current (2013) | Target (2050) | Growth required (% $yr^{-1}$) |
|---|---|---|---|---|
| $N_{ind}$-based B3 boundary | $Yield_{majorcrops}$ kgN $ha^{-1}$ | 47 | 70 | 1.29 |
| | $NUE_{crop}$ % | 38.5 | 70 | 0.83 |
| | Manure N recycling rate (β) % | 24 | 70 | 1.21 |
| | Livestock NCE % | 10 | 12.5 | 0.07 |
| | $r_{BNF}^{fixing}$ kgN $ha^{-1}$ | 15 | 15 | 0.00 |
| Organic B4b boundary | $Yield_{majorcrops}$ kgN $ha^{-1}$ | 38 | 57 | 1.36 |
| | $NUE_{crop}$ % | 38.5 | 70 | 0.83 |
| | Manure N recycling rate (β) % | 24 | 70 | 1.21 |
| | Livestock NCE % | 10 | 12.5 | 0.07 |
| | $r_{BNF}^{fixing}$ kgN $ha^{-1}$ | 15 | 25 | 1.75 |

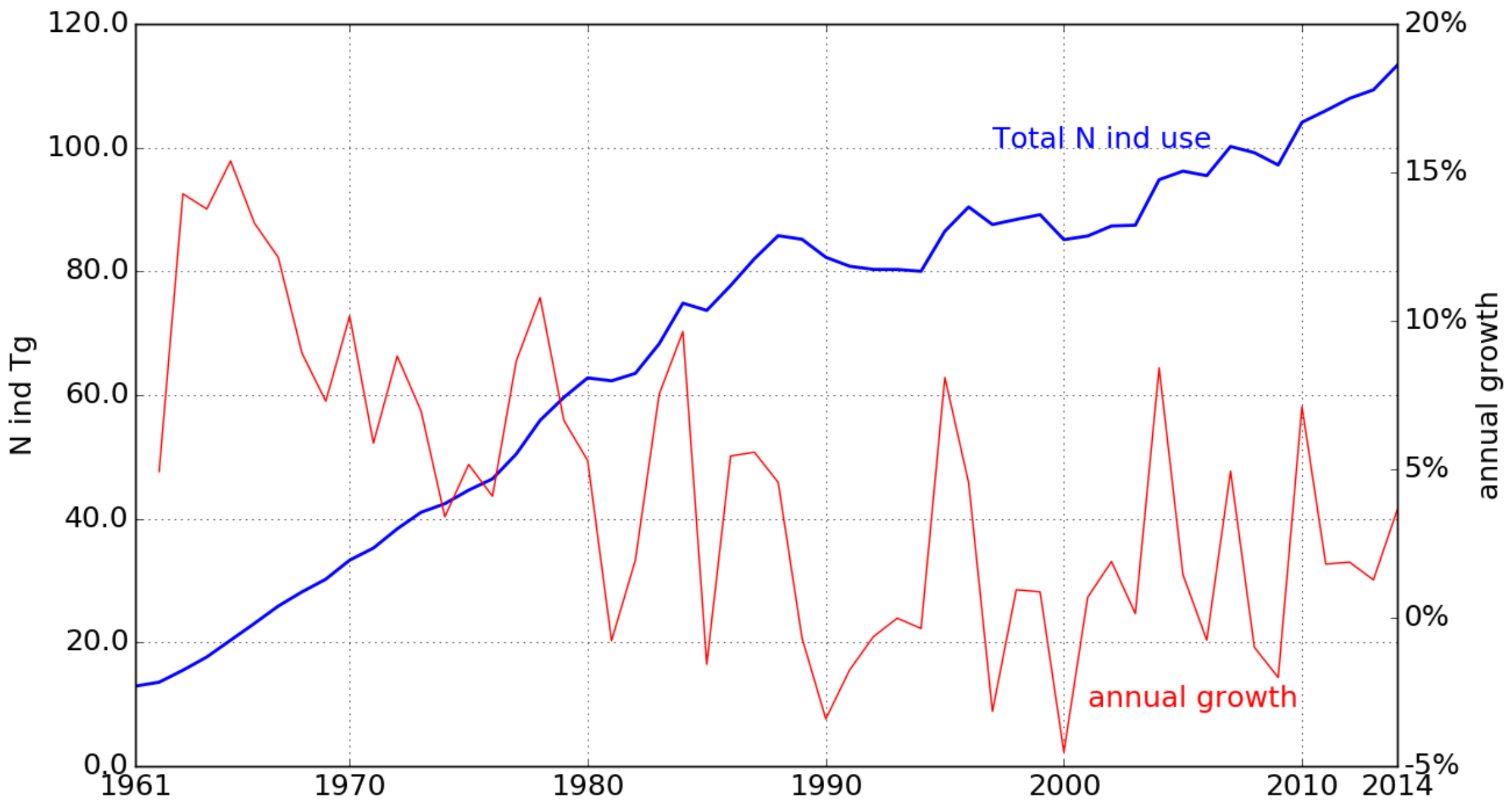

**Figure S11| Total $N_{ind}$ production and annual growth rate since 1961 (ref [2])**

## References in Supplementary Materials